\documentclass[lettersize,journal]{IEEEtran}
\usepackage{amsmath,amssymb,amsfonts}
\usepackage{algorithmic}
\usepackage{graphicx}
\usepackage{textcomp}
\usepackage{xcolor}
\usepackage{subfigure}

\usepackage{bm}
\usepackage{algorithm}
\usepackage{algorithmic}
\usepackage{soul}
\usepackage{balance}
\usepackage{cite}
\usepackage[version=3]{acro}

\DeclareAcronym{ICI}{short = ICI ,long  = inter-carrier interference ,tag = abbrev}
\DeclareAcronym{ISI}{short = ISI ,long  = inter-symbol interference ,tag = abbrev}
\DeclareAcronym{OFDM}{short = OFDM ,long  = orthogonal frequency division multiplexing ,tag = abbrev}
\DeclareAcronym{OTFS}{short = OTFS ,long  = orthogonal time frequency space ,tag = abbrev}
\DeclareAcronym{DFT}{short = DFT ,long  = discrete Fourier transform ,tag = abbrev}
\DeclareAcronym{IDFT}{short = IDFT ,long  = inverse discrete Fourier transform ,tag = abbrev}
\DeclareAcronym{ISFFT}{short = ISFFT ,long  = inverse symplectic finite Fourier transform ,tag = abbrev}
\DeclareAcronym{SFFT}{short = SFFT ,long  = symplectic finite Fourier transform ,tag = abbrev}
\DeclareAcronym{PAPR}{short = PAPR ,long  = peak-to-average power ratio ,tag = abbrev}
\DeclareAcronym{6G}{short = 6G ,long  = sixth generation ,tag = abbrev}
\DeclareAcronym{5G}{short = 5G ,long  = fifth generation ,tag = abbrev}
\DeclareAcronym{4G}{short = 4G ,long  = fourth generation ,tag = abbrev}
\DeclareAcronym{DD}{short = DD ,long  = delay-Doppler ,tag = abbrev}
\DeclareAcronym{TF}{short = TF ,long  = time-frequency ,tag = abbrev}
\DeclareAcronym{MIMO}{short = MIMO ,long  = multiple-input-multiple-output ,tag = abbrev}
\DeclareAcronym{3GPP}{short = 3GPP ,long  = 3rd generation partnership program ,tag = abbrev}
\DeclareAcronym{MA}{short = MA ,long  = multiple access ,tag = abbrev}
\DeclareAcronym{MUI}{short = MUI ,long  = multi-user interference ,tag = abbrev}
\DeclareAcronym{SINR}{short = SINR ,long  = signal to interference plus noise ratio ,tag = abbrev}
\DeclareAcronym{TDMA}{short = TDMA ,long  = time-division multiple access ,tag = abbrev}
\DeclareAcronym{FDMA}{short = FDMA ,long  = frequency-division multiple access ,tag = abbrev}
\DeclareAcronym{SNR}{short = SNR ,long  = signal to noise ratio ,tag = abbrev}
\DeclareAcronym{FCP}{short = FCP ,long  = full-\ac{CP},tag = abbrev}
\DeclareAcronym{HM}{short = HM ,long  = high mobility,tag = abbrev}
\DeclareAcronym{LM}{short = LM ,long  = low mobility,tag = abbrev}

\DeclareAcronym{AWGN}{short = AWGN ,long  =  additive white Gaussian noise,tag = abbrev}

\DeclareAcronym{BS}{short = BS ,long  =  base station,tag = abbrev}

\DeclareAcronym{DL}{short = DL ,long  =  downlink ,tag = abbrev}

\DeclareAcronym{UL}{short = UL ,long  =  uplink,tag = abbrev}
\DeclareAcronym{UE}{short = UE ,long  =  user equipment,tag = abbrev}
\DeclareAcronym{BER}{short = BER ,long  =  bit-error rate,tag = abbrev}

\DeclareAcronym{SC}{short = SC ,long  =  single-carrier,tag = abbrev}

\DeclareAcronym{BW}{short = BW ,long  =  band-width,tag = abbrev}

\DeclareAcronym{MSE}{short = MSE ,long  =  mean-square-error,tag = abbrev}

\DeclareAcronym{MU}{short = MU ,long  =  multi-user,tag = abbrev}

\DeclareAcronym{TD}{short = TD ,long  =  time delay,tag = abbrev}

\DeclareAcronym{ST}{short = ST ,long  =  symbol time,tag = abbrev}

\DeclareAcronym{SCS}{short = SCS ,long  =  sub-carrier spacing,tag = abbrev}

\DeclareAcronym{PEP}{short = PEP ,long  =  pairwise error probabilities,tag = abbrev}
\DeclareAcronym{BPSK}{short = BPSK ,long  =  binary phase shift keying,tag = abbrev}
\DeclareAcronym{QAM}{short = QAM ,long  =  quadrature amplitude modulation,tag = abbrev}

\DeclareAcronym{CP}{short = CP ,long  =  cyclic-prefix,tag = abbrev}

\DeclareAcronym{SDR}{short = SDR ,long  =  software-defined radio,tag = abbrev}

\DeclareAcronym{mMIMO}{short = mMIMO ,long  =  massive multiple input multiple output,tag = abbrev}

\DeclareAcronym{SIC}{short = SIC ,long  =  successive interference cancellation,tag = abbrev}

\DeclareAcronym{RIS}{short = RIS ,long  =  reconfigurable intelligent surfaces,tag = abbrev}

\DeclareAcronym{NOMA}{short = NOMA ,long  =  non-orthogonal MA,tag = abbrev}

\DeclareAcronym{RSMA}{short = RSMA ,long  =  rate splitting MA,tag = abbrev}

\DeclareAcronym{OFDMA}{short = OFDMA ,long  =  orthogonal frequency division multiple access,tag = abbrev}

\DeclareAcronym{MAC}{short = MAC ,long  =  medium access control,tag = abbrev}

\DeclareAcronym{INI}{short = INI ,long  =  inter-numerology interference,tag = abbrev}

\DeclareAcronym{iid}{short = i.i.d ,long  =  independent and identically distributed ,tag = abbrev}

\DeclareAcronym{CCDF}{short = CCDF ,long  =   complementary cumulative distribution function ,tag = abbrev}

\DeclareAcronym{PDP}{short = PDP ,long  =   power delay profile ,tag = abbrev}

\begin{document}

\title{A Novel Coexistence Scheme for OFDM and OTFS Waveforms for Uplink in 6G Networks}
  \author{ Badr Eddine Ouakouak, Salah Eddine Zegrar,~\IEEEmembership{Member,~IEEE}, and H\"{u}seyin Arslan,~\IEEEmembership{Fellow,~IEEE}

\thanks{Badr Eddine Ouakouak and H\"{u}seyin Arslan are with the Department of Electrical and Electronics Engineering, Istanbul Medipol University, Istanbul, 34810, Turkey (e-mail: badr.ouakouak@std.medipol.edu.tr; huseyinarslan@medipol.edu.tr).}
\thanks{Salah Eddine Zegrar is with KAUST, Thuwal, 23955, Saudi Arabia (e-mail: salaheddine.zegrar@kaust.edu.sa).}}

\maketitle

\begin{abstract}

The upcoming sixth generation (6G) wireless networks are designed to support advanced applications that the current air interface based on orthogonal frequency division multiplexing (OFDM) cannot accommodate, necessitating innovations in waveform design. Orthogonal time frequency space (OTFS) has emerged as a promising alternative, offering enhanced performance particularly in high-mobility channel environments.
However, deploying fundamentally different waveforms over the same frequencies fails to exploit the mutual relationships between them. 
In this paper, a forward-compatible yet backward-compatible coexistence framework is proposed, where  various OFDM waveform options are incorporated while seamlessly integrating OTFS within the same network.
The proposed coexistence paradigm involves utilizing a specially designed spreading matrix in the delay-Doppler (DD) domain to process OTFS data. The matrix is tailored to establish orthogonality between OTFS and OFDM waveforms, effectively reducing cross-interference when both coexist within the same time-frequency (TF) resources.
Our results demonstrate up to a 6 dB improvement in bit error rate (BER) performance for the OTFS waveform compared to other coexistence schemes. The proposed approach also shows greater interference resilience and approaches the upper bounds of achievable rates. By utilizing the full bandwidth and time duration, it enhances range and velocity resolutions, leading to improved sensing performance. These findings are supported by mathematical analysis and extensive simulations for uplink scenarios.
\end{abstract}


\section{Introduction} 
The vision for \ac{6G} encompasses not only higher data rates and lower latency but also advanced capabilities such as high-mobility communication, ultra-massive machine-type communication (umMTC), and extremely low-power communication \cite{zhang20196g}. Achieving these ambitious targets requires enhancements across all layers of the network stack, with the physical layer—particularly waveform design—being a critical area of focus \cite{yazar20206g}. In this context, \ac{OFDM} has been the waveform of choice for both \ac{4G} and \ac{5G} systems due to its robustness against multipath fading, high spectral efficiency, and ease of implementation. OFDM has demonstrated exceptional performance in satisfying a wide range of wireless communication requirements.

Despite its advantages, OFDM shows limitations in addressing the stringent demands of 6G networks \cite{10382693}. For instance, applications such as high-speed trains and high carrier frequencies, both of which are anticipated in 6G scenarios, expose OFDM's vulnerability to Doppler shifts and phase noise. These impairments lead to significant \ac{ICI}, which cannot be neglected \cite{wu2016survey}. To support high mobility and mitigate Doppler spread effects, 5G networks have enhanced conventional OFDM by adopting larger \ac{SCS} values, enabled through the concept of multi-numerology OFDM. While this approach improves robustness against channel variations, it also introduces challenges such as \ac{INI}, which can be further exacerbated by hardware impairments. Furthermore, OFDM's high \ac{PAPR} and large spectral sidelobes make it less suitable for energy-constrained devices, where constant-envelope signals are preferred for power efficiency. These properties also degrade sensing performance, particularly in low-power applications. These challenges collectively motivate the exploration of alternative waveforms capable of satisfying the broader set of 6G requirements.

 \Ac{OTFS} has emerged as a particularly promising option for 6G waveforms \cite{hadani2017orthogonal}. Unlike OFDM, which operates in the \ac{TF} domain, OTFS modulates information in the \ac{DD} domain, capturing the dynamics of time-varying wireless channels. This domain transformation renders OTFS highly resilient to Doppler shifts and multipath fading, making it particularly well-suited for \ac{HM} environments. OTFS also offers potential advantages in terms of reduced PAPR and enhanced performance in sparse channels \cite{surabhi2019peak}.

The design of \ac{MA} mechanisms is essential for enabling fully functional OTFS systems and has been the subject of various studies \cite{surabhi1902multiple,khammammetti2018otfs,augustine2019interleaved,correas2023mimo}. In \cite{surabhi1902multiple}, the authors evaluated the performance of OTFS-MA in \ac{HM} \ac{UL} scenarios using various \ac{DD} allocation strategies and compared it to \ac{OFDMA}, exploiting the benefits of functioning in  the DD domain. In \cite{khammammetti2018otfs}, methods for OTFS-MA in the TF domain were studied, where OTFS users occupy orthogonal portions of the TF resources. In \cite{augustine2019interleaved}, the authors demonstrated how interleaving users in the TF domain can greatly improve BER and PAPR of OTFS modulation. This work builds on this idea to design a scheme where OTFS users can coexist with OFDM, without undermining their performance. Similarly, in \cite{correas2023mimo}, the authors proposed arbitrary TF allocation of OTFS users to enhance sensing performance. In \cite{li2020new}, an OTFS-MA approach was developed that involves multiplexing UEs in space, using the angle of arrival/departure as a degree of freedom to separate users.

Although OTFS is well-suited for certain 6G scenarios, OFDM is expected to remain prevalent due to its existing infrastructure support and standardization. Replacing OFDM entirely with a new waveform is unlikely to be feasible given the diverse service requirements of 6G networks \cite{andrews20246}. This reinforces the need for a coexistence approach, wherein OTFS and OFDM operate together in the same system. Such a strategy enables leveraging the unique strengths of each waveform—deploying OTFS for scenarios requiring robustness and delay-Doppler sparsity, while retaining OFDM for compatibility and general-purpose applications. Integrating both waveforms within a unified design holds promise for improving overall network performance in heterogeneous conditions \cite{10382693}.

Although significant progress has been made toward fully capable OTFS-based networks \cite{surabhi1902multiple,khammammetti2018otfs,augustine2019interleaved,correas2023mimo}, ensuring seamless coexistence with legacy OFDM systems remains a key challenge. This highlights the critical need for further research into mechanisms and architectures that facilitate efficient integration of both waveforms in a single network.
One way to integrate the two waveforms is through a spatial scheduling paradigm, as proposed in \cite{chong2024hybrid}. In this scheme,
the two waveforms are split in space to meet the needs of users in a \ac{DL} \ac{mMIMO} system. Other studies \cite{ding2019otfs,zhang2022iterative} explore \ac{NOMA} techniques to allow users with different mobility profiles to share TF resources, employing \ac{SIC} for separation. 
For instance, in \cite{zhang2022iterative}, a NOMA-based coexistence of OTFS and OFDM users with heterogeneous mobility was investigated, where \ac{SIC} techniques were employed to achieve desired performance. A comparable method is employed in \cite{ge2021otfs}, where users with all mobility profiles utilize coded OTFS. These solutions produced excellent results, at the expense of complicated interference mitigation strategies. 
On the other hand, in \cite{lin2022resource}, groups of UEs are given orthogonal TF resources, according to their mobility characteristics. The authors of \cite{wu2022co} put out interesting plans for OTFS and OFDM users to exist in harmony in the TF domain; but, as this study will demonstrate, their performance deteriorates due to the restricted resources allotted to each user. Therefore, the issue of coexistence between OFDM and OTFS users still requires improved solutions. Other methods include \ac{MAC}-layer solutions involving \ac{RIS} \cite{cao2022massive}, or hybrid coexistence with \ac{RSMA} systems \cite{huai2024cross}. The last two approaches suffer particularly from increased latency despite their remarkable success in other aspects.

While the coexistence of OTFS and OFDM UEs can be managed by assigning orthogonal TF resources, this approach compromises a core advantage of OTFS, i.e., sparsity in the DD domain. Reserving a portion of the TF grid for OTFS signaling significantly diminishes its performance, as OTFS relies on spreading signals over extended \ac{ST} and utilizing a wide \ac{BW} to fully exploit the benefits of DD-domain communication. This, in turn, puts OTFS at a disadvantage compared to OFDM.

To fully leverage the benefits of \Ac{OTFS} while minimizing the interference with \Ac{OFDM}, in this paper, we propose a novel coexistence scheme that involves \Ac{DD} spreading. Inspired by multiple access techniques in both waveforms, the proposed approach consists of applying a specifically designed spreading matrix in the \Ac{DD} domain to the \Ac{OTFS} data. The spreading matrix is optimized to create orthogonality between the \Ac{OTFS} and \Ac{OFDM} waveforms, thereby minimizing the cross-interference when both waveforms coexist within the same frequency band. The DD spreading technique enhances the resilience of OTFS against interference from OFDM by ensuring that the power of the OTFS signal is concentrated in specific TF bins that do not overlap with the resources occupied by OFDM. This selective spreading not only preserves the full power and robustness of OTFS in challenging conditions but also ensures that the coexistence with OFDM does not lead to significant degradation in system performance. 

Our contributions are summarized as follows: 
\begin{enumerate}
    \item We propose a novel coexistence scheme where OTFS and OFDM waveforms are scheduled in a way that maximum gain is achieved by both OTFS and OFDM users. By employing a spreading matrix in the DD domain, users are scheduled in the TF domain in a manner that controls \ac{MUI} and maximizes resource utilization. 
    \item We derive analytical expressions for the interference experienced by each user under different configurations of the spreading parameters. Then. we further analyze the upper bounds on achievable rates, demonstrating that the proposed system offers more robust and tighter bounds, and significant flexibility in terms of interference management. This flexibility is critical for accommodating diverse service requirements and channel conditions in multi-user, multi-waveform systems.
    \item We demonstrate that the proposed coexistence scheme is capable of resolving a higher number of channel taps and leveraging full channel diversity, which is particularly advantageous in rich scattering environments. Through theoretical analysis, we derive lower bounds for the \ac{BER} under special cases and highlight the advantages of the proposed coexistence scheme. 
    \item We demonstrate that the proposed system enhances range and velocity resolutions through spreading by efficiently utilizing the full bandwidth and time duration for OTFS users. Analytical results highlight the system's ability to achieve improved sensing performance\footnote{\textit{Notations:} Bold uppercase $\mathbf{A}$, bold lowercase $\mathbf{a}$, and non-bold letters $A,a$ denote matrices, column vectors, and scalar values, respectively. $(\cdot)^\dagger$, $(\cdot)^T$, and $(\cdot)^{-1}$ denote the Hermitian, transpose, and inverse operators. $\text{vec}\{\cdot\}$ takes the elements of a matrix column-wise and arranges them in a single column vector. $\text{Tr}\{\cdot\}$ and $\text{det}\{\cdot\}$ denote the trace and determinant operations, respectively. $[\alpha]_{\beta}$ takes modulo-$\beta$ of $\alpha$. $\delta_n = \delta(n)$ denotes the Dirac-delta function and $\operatorname{E}(\cdot)$ denotes the expectation operator. $\operatorname{diag}\left( {A}_1,\dots,{A}_{N} \right)$ denotes the diagonal matrix composed of ${A}_1,\dots,{A}_{N}$ on its diagonal. $\mathbb{C}^{{M\times N}}$ denotes the space of $M\times N$ complex-valued matrices. $\mathbf{A} \otimes \mathbf{B}$ is the Kronecker product of $\mathbf{A}$ and $\mathbf{B}$ and $\mathbf{0}_{N}$ is the matrix of size $N \times N$ with all elements being zeros. When not used as an index, symbol $j$ represents the imaginary unit of complex numbers with $j^2=-1$.}.
\end{enumerate}  
The remainder of this article is organized as follows;
Section II provides the signal model of both OTFS and OFDM at the transmitter and receiver sides. In Section III, we present the proposed OFDM-OTFS coexistence scheme. Section IV evaluates and analyzes the performance of the proposed scheme in terms of achievable rates, \ac{BER}, and sensing capabilities. Section V then presents numerical results for validation. The concluding remarks are given in Section VI.


\section{Preliminaries}

Consider a wireless communication system where $M\times N$ data symbols are to be transmitted over a total \ac{BW} of $B$ centered at the carrier frequency $f_c$. 

\subsection{OFDM Signal Model}
The \ac{OFDM} modulator distributes the symbols $\{X_{\mathrm{TF}}^{\mathrm{OFDM}}(m,n)\}$ at $\{m=0,\dots,M-1\}$ frequency subcarriers at each time $\{n=0,\dots,N-1\}$. The subcarrier spacing is given by $\Delta f= B/M$ and the time duration of one \Ac{OFDM} symbol is $T = 1/\Delta f$. The frequency samples $X_{\mathrm{TF}}^{\mathrm{OFDM}}(m,n)$ need to be converted to time domain using the \ac{IDFT} operation. Then, the transmitted time-domain \Ac{OFDM} signal becomes \cite{banelli2014modulation}
 \begin{equation} \small
s_{\mathrm{OFDM}}(p)=\sum_{n=0}^{N-1}\sum_{m=0}^{M-1}  X_{\mathrm{TF}}^{\mathrm{OFDM}}(n,m) e^{\frac{j 2 \pi m p}{M}}  \operatorname{rect}\big(\frac{p-nN}{N}\big),
\label{equ:tx_signal_sample_form}
\end{equation}
where rect$(p) = G_\text{Heaviside}(p) - G_\text{Heaviside}(p - 1)$ is the rectangular pulse function and $G_\text{Heaviside}(p)$ is the Heaviside step function. The matrix form of \eqref{equ:tx_signal_sample_form} is given as
\begin{equation}
    \mathbf{s}_{\mathrm{OFDM}}= \text{vec}\{\mathbf{F}_{M}^{\dagger} \mathbf{X}_{\mathrm{TF}}^{\mathrm{OFDM}}\} = (\mathbf{I}_{N}\otimes \mathbf{F}^\dagger_M)\mathbf{x}^{\mathrm{OFDM}}_{\mathrm{TF}},
    \label{equ:TX_signal_proposed1}
\end{equation}
where $\mathbf{F}_{N}$ is the $N\times N$ \ac{DFT} matrix and $\mathbf{x}^{\mathrm{OFDM}}_{\mathrm{TF}} = \text{vec}\{\mathbf{X}_{\mathrm{TF}}^{\mathrm{OFDM}}\}$.

\subsection{OTFS Signal Model}
 
First, the \Ac{OTFS} modulator distributes $\{X_{\mathrm{DD}}(k,l),~k=0,\dots,N-1, ~ l = 0,\dots,M-1\}$ symbols in the two-dimensional \Ac{DD} grid. Then, the samples $X_{\mathrm{DD}}(k,l)$ are converted to the \Ac{TF} domain grid $\{X_{\mathrm{TF}}^{\mathrm{OTFS}}(n,m),~n=0,\dots,N-1, ~ m = 0,\dots,M-1\}$ through the \ac{ISFFT} as follows: 
\begin{equation}
    X_{\mathrm{TF}}^{\mathrm{OTFS}}(n, m)=\frac{1}{\sqrt{MN}} \sum_{l=0}^{M-1} \sum_{k=0}^{N-1} X_{\mathrm{DD}}(k,l) e^{j 2 \pi\left(\frac{n k}{N}-\frac{m l}{M}\right)}.
    \label{equ:ISFFT}
\end{equation}
The \Ac{TF} signal ${X}_{\mathrm{TF}}^{\mathrm{OTFS}}$ in \eqref{equ:ISFFT} needs to be converted to time domain before transmission. For this purpose, practical \ac{OFDM} systems, which are based on the \ac{IDFT}, are used. Hence, the discrete output of an OFDM modulator is 
\begin{equation}\small
s_{\mathrm{OTFS}}(p)=\sum_{n=0}^{N-1}\sum_{m=0}^{M-1}  {X}_{\mathrm{TF}}^{\mathrm{OTFS}}(n,m) e^{\frac{j 2 \pi m p}{M}}  \operatorname{rect}\left(\frac{p-mN}{N}\right).
\label{equ:dd}
\end{equation}
The expression \eqref{equ:dd} is written in matrix form as follows \cite{raviteja2018practical}:
\begin{equation}
    \mathbf{X}_{\mathrm{TF}}^{\mathrm{OTFS}} = \mathbf{F}_{M}\mathbf{X}_{\mathrm{DD}}\mathbf{F}_{N}^{\dagger},
\end{equation}
and
\begin{equation}
     \mathbf{x}_{\mathrm{OTFS}}= (\mathbf{F}^\dagger_N\otimes\mathbf{F}_{M})\mathbf{x}_{\mathrm{DD}},
\end{equation}
where $\mathbf{x}_{\mathrm{DD}} = \text{vec}\{\mathbf{X}_{\mathrm{DD}}\}$, and $\mathbf{x}_{\mathrm{OTFS}} = \text{vec}\{\mathbf{X}_{\mathrm{TF}}^{\mathrm{OTFS}}\}$.

\subsection{The Channel Model}

To achieve coexistence between the two waveforms, the OFDM and OTFS frames have to be aligned in the TF domain, to avoid possible \ac{ISI} issues. To this end, full-\ac{CP} \ac{OTFS} configuration, with a \ac{CP} length of $L_{cp}$, is considered in this scheme. The resulting time-domain doubly dispersive channel matrix is given by \cite{zegrar2022effect}
\begin{equation}
    \hat{\mathbf{H}} =\sum_{i=1}^{P} h_i \boldsymbol{\Pi}^{l_i}_{(M+L_{cp})N}\boldsymbol{\Omega}^{k_i}_{(M+L_{cp})N},
    \label{equ:time_domain_channel}
\end{equation}
with $\boldsymbol{\Pi}_{\eta}$ being the $\eta\times \eta$ permutation matrix defined as 
\begin{equation}
    \boldsymbol{\Pi}_{\eta}^{l_i}(m',n') = \delta([m'-n'-l_i]_{\eta}),~\text{for}~ m',n'\in [0,\eta-1],
\end{equation} 
and $\boldsymbol{\Omega}_{\eta}$ being an $\eta\times \eta$ diagonal matrix, given by
\begin{equation}
\boldsymbol{\Omega}^{k_i}_{\eta}(m',n')= e^{j\frac{2\pi m' k_i}{\eta}} \delta(m'-n').
\end{equation}
Here, the Doppler $k_i$ contains both the integer and fractional parts. In contrast, delays are assumed to be integer-valued in this work. Since user separation is achieved in the TF domain, fractional delay components can generally be tolerated without significant degradation. However, if fractional delays cannot be neglected, transmitter-side windowing may be applied to minimize the resulting interference. Additionally, assuming integer delays enables the derivation of closed-form expressions for the achievable rates and \ac{BER} performance.

It can be shown that after discarding the CP, the time-domain channel with full-CP  can be represented by a block diagonal matrix given by \cite{zegrar2022effect}
\begin{equation}
    \mathbf{H}_\text{T} = \text{diag}\left (\mathbf{H}_0,...,\mathbf{H}_{N-1}  \right),
    \label{equ:Ht}
\end{equation}
where
\begin{equation}
    \mathbf{H}_{n} =\sum_{i=1}^{P} h_i \boldsymbol{\Pi}^{l_i}_{M}\boldsymbol{\Omega}^{k_i,n}_{M},
\end{equation}
where $\boldsymbol{\Omega}^{k_i,n}_{M}(m',n') = \boldsymbol{\Omega}^{k_i}_{M}(m',n') ~\text{for}~m'\in [n(M+L_{\text{cp}})+L_{\text{cp}},(n+1)(M+L_{\text{cp}})-1]$.
Then, the \Ac{TF} effective channel can be obtained by
\begin{equation}
   \mathbf{H}_{\text{TF}} =(\mathbf{I}_N\otimes \mathbf{F}_M)\mathbf{H}_{\text{T}}( \mathbf{I}_N \otimes \mathbf{F}^\dagger_M).
   \label{equ:Htf}
\end{equation}
{\textbf{Lemma 1}}: $ \mathbf{H}_{\text{TF}} $ is also block diagonal given by
\begin{equation}
    \mathbf{H}_\text{TF} = \text{diag}\big (\hat{\mathbf{H}}_0,...,\hat{\mathbf{H}}_{N-1}  \big),
  \label{equ:H_tf}
\end{equation}
where each block is given by
\begin{equation}
    \hat{\mathbf{H}}_n = \mathbf{F}_M\mathbf{H}_{n} \mathbf{F}^\dagger_M.
    \label{equ:momo}
\end{equation}
\textbf{Proof}: let $\mathbf{E}_n$ be the $n$-th $N\times N$ basis matrix i.e., $\mathbf{E}_n$ is a diagonal matrix with all entries being zero except for the $n$-th diagonal element being one, i.e.,
\begin{equation}
    \mathbf{E}_n(m',n') = \delta(n'-n)\delta(m'-n).
\end{equation}
Then, \eqref{equ:Ht} can be rewritten as follows:
\begin{equation}
    \mathbf{H}_\text{T} = \sum_{n=0}^{N-1}\mathbf{E}_n\otimes \mathbf{H}_n .
    \label{equ:jojo}
\end{equation}
Substituting \eqref{equ:jojo} in \eqref{equ:Htf}, and
using the mixed-product property, we find
\begin{equation}
\begin{aligned}
    \mathbf{H}_{\text{TF}} = \sum_{n=0}^{N-1}\mathbf{E}_n\otimes( \mathbf{F}_M\mathbf{H}_n \mathbf{F}^\dagger_M),
        \label{equ:TF koko}
\end{aligned}
\end{equation} 
which shows that it is indeed block diagonal, with the diagonal block given by \eqref{equ:momo}.\hfill$\blacksquare$

\subsection{Receiver}
 The received TF-domain signal can be expressed using the input-output relation \cite{raviteja2018practical} given as
 \begin{equation}
 \mathbf{y}=\mathbf{H}_{\mathrm{eff}} \mathbf{x}+\mathbf{w},
 \label{equ:y}
\end{equation}
where $\mathbf{w}$ is the \ac{AWGN} vector and $\mathbf{H}_{\mathrm{eff}} \in \mathbb{C}^{M N \times M N}$ denotes the effective channel which is expressed by $\mathbf{H}_{\mathrm{eff}} = (\mathbf{I}_N\otimes\mathbf{F}_M  )\mathbf{H}( \mathbf{I}_N\otimes \mathbf{F}_M^{\dagger})$ and $\mathbf{H}_{\mathrm{eff}} = (\mathbf{F}_N \otimes \mathbf{I}_M)\mathbf{H}(\mathbf{F}_N^{\dagger} \otimes \mathbf{I}_M)$ for OFDM and OTFS \acp{UE}, respectively.

\section{Proposed OFDM-OTFS Coexistence Scheme}

An \ac{UL} scenario involving one \ac{BS} and $U$ UEs is considered in the system model, where the BS serves multiple UEs using different PHY technologies, namely, OFDM and OTFS signaling\footnote{Although the analysis is presented for conventional CP-OFDM, the proposed scheme is compatible with low-PAPR waveforms such as SC-FDMA and DFT-s-OFDM, which are widely used in practical uplink scenarios. These waveforms, which are significant for the large-scale deployment of the coexistence system, can be incorporated with minor transmitter-side modifications.}. Based on the \ac{UE}'s condition and requirements, the \ac{BS} decides what waveform to use\footnote{An example of waveform assignment to UEs can be based on their mobility profiles, where low-mobility UEs are served using OFDM, while high-mobility UEs are served using OTFS.}.\\
The \ac{TF}-domain signal at the transmitter corresponding to the $j$-th user can be written as
\begin{equation}
     \mathbf{s}^{(j)}_{\text{TF}} =\mathbf{Q}^{(j)}\mathbf{x}^{(j)} ,
   \label{equ:mld}
\end{equation}
where $\mathbf{x}^{(j)} = \text{vec}\{\mathbf{X}^{(j)}_{\mathrm{TF}}\}$ denotes the TF domain samples in vector form and $\mathbf{Q}^{(j)}$ is the allocation matrix of the $j$-th \ac{UE}. 
The allocation matrix $\mathbf{Q}^{(j)}$ will be defined and studied in the next section, and it will mainly serve as a mathematical representation of the spreading of the \ac{TF} samples corresponding to the $j$-th user. This formulation is general and allows for both OFDM and OTFS-based transmissions under a unified framework. Using \eqref{equ:TX_signal_proposed1}, the transmitted time domain signal of the $j$-th user can be written as follows:

\begin{equation}
    \mathbf{s}^{(j)}_{\text{tx}} =(\mathbf{I}_{N}\otimes \mathbf{F}^\dagger_M)\mathbf{s}^{(j)}_{\text{TF}}.
\end{equation}

At the receiver side, the signals from all transmitters are superimposed. After discarding the CP, the received \ac{TF}-domain signal at the \ac{BS} is given by
\begin{equation}
    \mathbf{y} =\sum_{j=1}^{U}\mathbf{H}^{(j)}_{\text{TF}}\mathbf{Q}^{(j)}\mathbf{x}^{(j)} + \mathbf{w}.
    \label{equ:in_out_ul}
\end{equation} 

\begin{figure}[t]
\centering
\includegraphics[scale = .3]{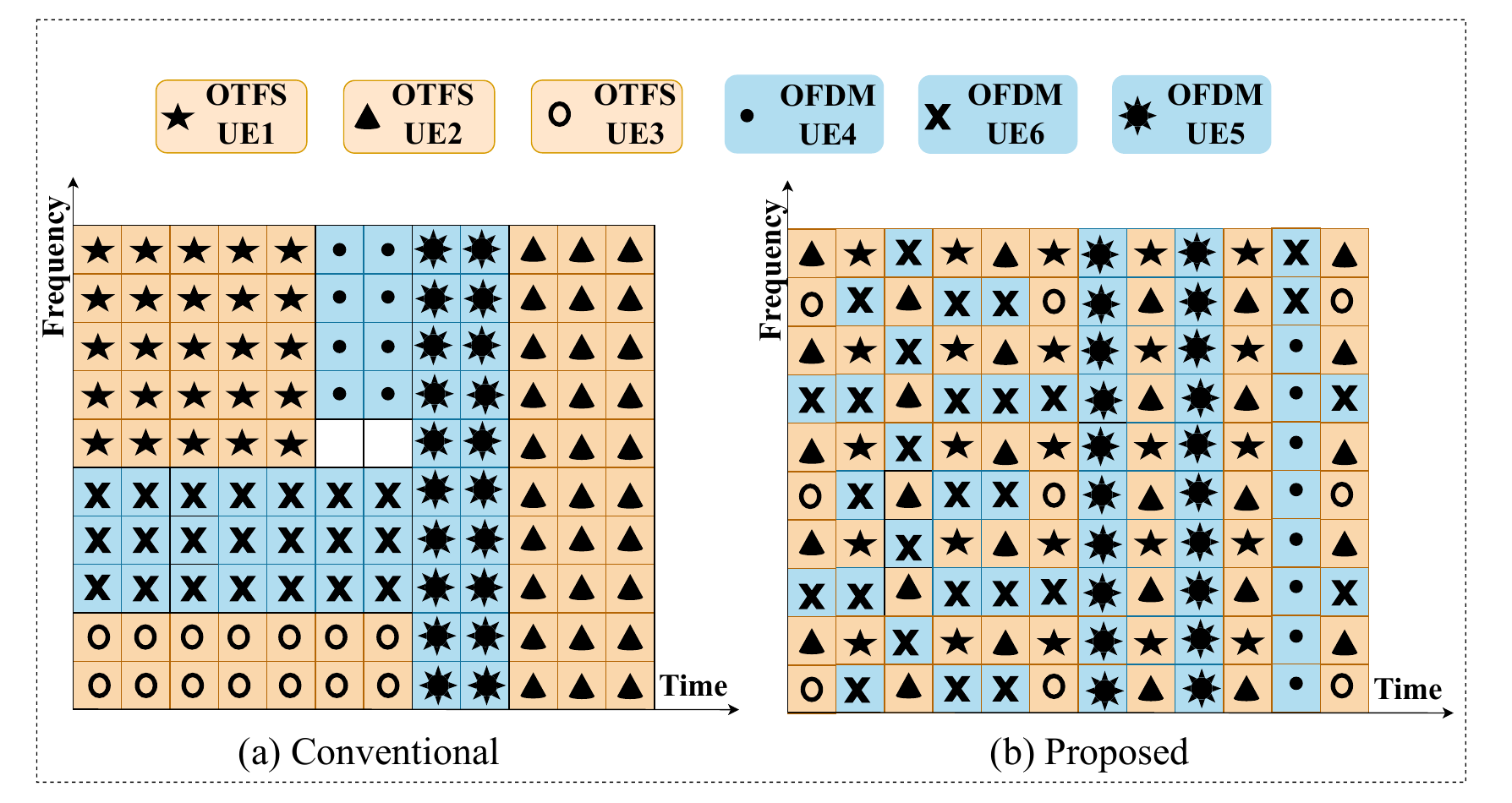}
\caption{Illustration of how the proposed scheme spreads the samples in the \Ac{TF} domain.}
\label{fig:proposed_spreading_illustration}
\end{figure} 

\subsection{Problem Formulation and Assumptions}
Figure~\ref{fig:proposed_spreading_illustration} illustrates the key distinction between prior coexistence schemes and the proposed method. Given the same total TF resources for the same number of users, the proposed scheme aims to allow each user to utilize as much as possible of these resources by spreading the samples to occupy the entire \ac{BW} and/or time duration. This spreading is governed by specific parameters, which will be defined and analyzed later in this work. The rationale for this is that ensuring TF orthogonality when scheduling UEs in the \ac{TF} domain when employing \ac{OFDM} and \ac{OTFS} modulations, as depicted in Fig.~\ref{fig:proposed_spreading_illustration}(a), can severely constrain OTFS's performance. Limiting the TF extent allocated to OTFS undermines one of its core advantages: its fine resolution in the DD domain. Such constraints reduce the diversity and robustness OTFS is designed to offer. Furthermore, the proposed spreading strategy provides additional benefits, which will be supported analytically and demonstrated through simulations in subsequent sections.

To address these limitations, this paper adopts a general waveform framework in which rectangular pulse shaping is applied during modulation. This can be represented by  
\begin{align}
\label{eq:6G_Waveform}
    x(t) &= \sum_{n =0 }^{N-1} \sum_{m =0}^{M-1} \underbrace{f(X_{n,m})}_\text{Precoding}  
   \overbrace{\text{rect}\left(\frac{t - nT}{T}\right)}^\text{Uniform Pulse Shaping}    e^{j 2\pi \Delta f n m}.
\end{align}
Furthermore, this paper proposes an innovative resource allocation scheme tailored for \ac{MU} scenarios involving different waveforms, specifically OFDM and OTFS modulation. The proposed approach ensures that the OTFS signal can fully exploit the available TF resources, thereby preserving its maximum DD resolution, while simultaneously minimizing or completely avoiding interference with OFDM signals in the TF domain. 
While OTFS operates natively in the DD domain, TF domain control is achieved through DD domain processing, enabled by the application of a spreading matrix $\mathbf{\Psi}_{a,b}$. The assumptions of the proposed framework are summarized as
\begin{itemize}
    \item OTFS employs rectangular pulse shaping.
    \item Since OFDM data is defined in the TF domain, resource allocation is straightforward, i.e., data symbols are mapped directly to the intended resources.
    \item To achieve optimal performance with OTFS, DD domain data is spread across the TF grid while minimizing the interference between OTFS and OFDM UEs, i.e., $f(\mathbf{X}_{\text{DD}}^\text{OTFS}(n,m)) = \text{ISFFT}\left\{\mathbf{\Psi}_{a,b} \otimes \mathbf{X}_{\text{DD}}^\text{OTFS}\right\}$.
    \item To avoid \ac{ISI} between successive symbols, a full-CP OTFS configuration is adopted.
\end{itemize}

\subsection{Proposed Delay-Doppler Spreading}
This subsection details the design principles and operational mechanics of $\mathbf{\Psi}_{a,b}$, highlighting how it facilitates resource orthogonality and interference avoidance. The  matrix $\mathbf{\Psi}_{a,b}$ is defined as follows:
\begin{equation}
    \mathbf{\Psi}_{a,b} = \mathbf{f}_{\alpha,a}\otimes\mathbf{f}^\dagger_{\beta ,b},
\end{equation}
where $\mathbf{f}_{m',n'}$ denotes the $n'$-th row of the $m'\times m'$ \ac{DFT} matrix. The parameters $a = 0,1,\dots,\alpha-1$ and $b = 0,1,\dots,\beta-1$ determine the starting positions in time and frequency, respectively, while $M' = M/\beta \in \mathbb{N}$ and $N' = N/\alpha \in \mathbb{N}$ define the grid granularity. The parameters $\alpha$ and $\beta$ determine the spreading factor in time and frequency, respectively.

Using $\mathbf{\Psi}_{a,b}$, the user's \ac{DD} domain samples are spread as given by
\begin{equation}
    \mathbf{S}_{\text{DD}}^\text{OTFS} \overset{\Delta}{=} \mathbf{\Psi}_{a,b}\otimes \mathbf{X}_{\text{DD}}^\text{OTFS}.
    \label{equ:good_day}
\end{equation}
{\textbf{Theorem 1} }: The resulting \ac{TF} samples achieve the desired resource allocation based on the parameters $\alpha$, $\beta$, $a$, and $b$ as in \eqref{equ:spread_theorem}.

\textbf{Proof}: The proof is directly obtained from the ISFFT of $\mathbf{S}_{\text{DD}}^\text{OTFS}$. We start by writing its expression
\begin{equation}
    \text{S}_{\text{DD}}^\text{OTFS}(k,l) =\frac{1}{\sqrt{\alpha \beta}}  X_{\mathrm{DD}}^{\mathrm{OTFS}}(k^\prime, l^\prime )e^{-j 2 \pi\left(\frac{v a}{\alpha}-\frac{u b}{\beta}\right)}.
    \label{equ:recov}
\end{equation} 
Let $ k = k^\prime + vN^\prime $ and $ l = l^\prime + uM^\prime $, with $k^\prime = 0,\dots,N^\prime-1$ , $l^\prime = 0,\dots,M^\prime-1$, $v = 0,\dots,\alpha-1$, and $u = 0,\dots,\beta-1$. Then, the \Ac{TF} signal is given by
\begin{equation}
   \text{S}_{\mathrm{TF}}^{\mathrm{OTFS}}(n, m)=\frac{1}{\sqrt{M N}} \sum_{l=0}^{M-1} \sum_{k=0}^{N-1} \mathbf{S}_{\text{DD}}^\text{OTFS}(k,l) e^{j 2 \pi\left(\frac{n k}{N}-\frac{m l}{M}\right)}.
   \label{equ:Stf}
\end{equation}
It can then be shown that \eqref{equ:Stf} simplifies to 
\begin{equation}
    S_{\mathrm{TF}}^{\mathrm{OTFS}}(n, m)=
    \frac{\text{T}(n, m)}{\alpha \beta} \sum_{u=0}^{\beta-1}\sum_{v=0}^{\alpha-1}e^{j 2 \pi\left(\frac{(n-a)v}{\alpha}-\frac{(m-b)u}{\beta}\right)},
    \label{eq_for_T}
\end{equation}
where $\text{T}(n, m)=\text{ISFFT}\left\{\mathbf{S}_{\text{DD}}^\text{OTFS}(k,l)e^{j 2 \pi\left(\frac{a k }{N}-\frac{b l }{M}\right)}\right\}$.\\
Observe that this sum is either 0 or $\alpha\beta$ depending on the values of $n$ and $m$. So, we find
\begin{equation}
    \text{S}_{\text{TF}}^\text{OTFS}(n,m) =
\begin{cases} 
    \text{T}(n,m)&,~n = n^\prime\alpha +a,~m=m^\prime\beta+b\\
    0 &, \text{otherwise} ,
\end{cases}
    \label{equ:spread_theorem}
\end{equation}
where $n^\prime = 0,\dots,N^\prime-1$, and $m^\prime = 0,\dots,M^\prime-1$.\\

This result confirms that the TF-domain resource allocation of the OTFS UE is governed by the parameters $\alpha$, $\beta$, $a$, and $b$. Specifically, by pre-processing the original DD-domain data symbols with the spreading matrix $\mathbf{\Psi}_{a,b}$, the resulting TF-domain signal is spread by a factor of $\alpha$ along the time axis starting at index $a$, and by a factor of $\beta$ along the frequency axis starting at index $b$. This completes the proof\footnote{The \ac{DD}-domain data symbols for a given user can be perfectly recovered from the spread \ac{TF}-domain samples, provided that the spreading matrix $\mathbf{\Psi}_{a,b}$ is known at the receiver.}.\hfill $\blacksquare$


\subsection{\Ac{TF} Domain Multiplexing of OTFS and OFDM UEs}

To evaluate the proposed OFDM–OTFS coexistence scheme, consider a scenario where $U \leq \alpha\beta$ users are multiplexed, consisting of both \ac{OFDM} and \ac{OTFS} UEs. Let $(a^{(k)},b^{(k)})$ be the corresponding starting indices for the $k$-th user, $k = 1,..,U$. The mapping between $k$ values and pairs $(a,b)$ is irrelevant, it merely has to be one-to-one\footnote{Here, $\alpha$ denotes the number of users that can be multiplexed in time, and $\beta$ the number of users that can be multiplexed in frequency. Two special cases arise when $\alpha = 1$ or $\beta = 1$, corresponding to \ac{FDMA} and \ac{TDMA} configurations, respectively.}.
Let $\mathbf{x}^{(j)} \in \mathbb{C}^{M^\prime N^\prime \times 1}$ be the TF domain samples in vector form of the $j$-th user, this user being defined by $(a^{(j)},b^{(j)})$. If user $j$ employs OFDM, then $\mathbf{x}^{(j)}$ directly represents its modulated data symbols. If user $j$ employs OTFS, then $\mathbf{x}^{(j)}$ corresponds to the values of $\text{T}(n', m')$ defined in~\eqref{equ:spread_theorem}.

Let the allocation matrix $\mathbf{Q}^{(j)}$ for the $j$-th user serve as the mathematical representation of the proposed TF-domain resource allocation scheme. To multiplex the different users in the TF domain, we define the allocation matrix as
\begin{equation}
    \mathbf{Q}^{(j)} = (\mathbf{I}_{N^\prime}\otimes \mathbf{a}^{(j)})\otimes(\mathbf{I}_{M^\prime}\otimes \mathbf{b}^{(j)}),
    \label{equ:ind}
\end{equation}
where  $\mathbf{a}^{(j)}$ and $\mathbf{b}^{(j)}$ are the $a^{(j)}$-th and $b ^{(j)}$-th standard basis vector in $\mathbb{R}^\alpha$ and $\mathbb{R}^\beta$, respectively. That is,  $\mathbf{a}^{(j)}$ and $\mathbf{b}^{(j)}$ have $1$ at the $a^{(j)}$-th and $b ^{(j)}$-th entry, respectively, and zeros otherwise.

{\textbf{Lemma 2}}:
The allocation matrices for the $k$-th user and the $j$-th user satisfy
\begin{equation}
    (\mathbf{Q}^{(k)})^\dagger\mathbf{Q}^{(j)} = \delta_{k,j} \mathbf{I}_{M^\prime N^\prime}.
\end{equation}
\textbf{Proof}:
First, we know that ${(\mathbf{a}^{(k)})}^{\dagger}\mathbf{a}^{(j)} = {(\mathbf{b}^{(k)})}^\dagger\mathbf{b}^{(j)} = \delta_{k,j}$. Then, 
\begin{equation}
    \begin{aligned}
 (\mathbf{Q}^{(k)})^\dagger\mathbf{Q}^{(j)} =&  
 {\left[(\mathbf{I}_{N^\prime}\otimes \mathbf{a}^{(k)})\otimes(\mathbf{I}_{M^\prime}\otimes \mathbf{b}^{(k)})\right]}^\dagger\\
          &\times\left[(\mathbf{I}_{N^\prime}\otimes \mathbf{a}^{(j)})\otimes(\mathbf{I}_{M^\prime}\otimes \mathbf{b}^{(j)})\right].
        \label{equ:fofo}
\end{aligned}
\end{equation}
Using the mixed-product property twice\cite{schacke2004kronecker}, \eqref{equ:fofo} simplifies to
\begin{equation}
    \begin{aligned}
 (\mathbf{Q}^{(k)})^\dagger\mathbf{Q}^{(j)} &=  \big[\mathbf{I}_{N^\prime}\otimes {(\mathbf{a}^{(k)})}^\dagger\mathbf{a}^{(j)}\big]\otimes \big[\mathbf{I}_{M^\prime}\otimes {(\mathbf{b}^{(k)})}^\dagger \mathbf{b}^{(j)}\big]\\ 
          &= \delta_{k,j} \mathbf{I}_{M^\prime N^\prime}.\\
\end{aligned}
\end{equation}
This completes the proof. \hfill$\blacksquare$

The matrix $(\mathbf{Q}^{(k)})^\dagger$ plays an important role at the receiver in the proposed scheme for detecting each UE's data. While the allocation matrices are orthogonal by design prior to transmission, the propagation channel can disturb this orthogonality introducing interference. Therefore, the effect of the \ac{TF} channel needs to be studied\footnote{Throughout, we will refer to the duration of one OFDM symbol as a ``time index'' or ``time resource''. Similarly, we refer to a frequency band of one \ac{SCS} as a ``frequency index '' or ``frequency resource''.}.

{\textbf{Lemma 3}}: 
Let $\mathbf{H}_{\text{TF}}$ be an $MN \times MN$ block-diagonal channel matrix. For two users $k$ and $j$ that occupy different time indices, i.e., $a^{(k)} \neq a^{(j)}$, we have
\begin{equation}
{(\mathbf{Q}^{(k)})}^\dagger\mathbf{H}_{\text{TF}}\mathbf{Q}^{(j)} = \mathbf{0}_{M^\prime N^\prime}.
\end{equation}
\textbf{Proof}: First, note that $\mathbf{H}_{\text{TF}}$  can be expressed as 
\begin{equation}
    \mathbf{H}_{\text{TF}} = \sum_{i=0}^{N^\prime-1}\sum_{t=0}^{\alpha-1}({\mathbf{E}}_i\otimes \mathbf{E}_t)\otimes\hat{\mathbf{H}}_{\alpha i + t},
\end{equation}
where $\mathbf{E}_i$ is the $i$-th basis matrix, see \textbf{Lemma 1}; and $\hat{\mathbf{H}}_i$ is an arbitrary $M \times M$ matrix. Then,
\begin{equation}
{(\mathbf{Q}^{(k)})}^\dagger\mathbf{H}_{\text{TF}}\mathbf{Q}^{(j)} = \sum_{i=0}^{N^\prime-1}\sum_{t=0}^{\alpha-1}{(\mathbf{Q}^{(k)})}({\mathbf{E}}_i\otimes \mathbf{E}_t)\otimes\hat{\mathbf{H}}_{\alpha i + t}\mathbf{Q}^{(j)}.
\label{equ:toto}
\end{equation}
Using the mixed-product property,  we can simplify \eqref{equ:toto} as 
\begin{equation}
{(\mathbf{Q}^{(k)})}^\dagger\mathbf{H}_{\text{TF}}\mathbf{Q}^{(j)} =
    \sum_{i=0}^{N^\prime-1}\sum_{t=0}^{\alpha-1}{\mathbf{A}}_{\alpha i + t} \otimes  {\mathbf{H}^\prime}_{\alpha i + t},
    \label{equ:summ}
\end{equation}
where ${\mathbf{A}}_{\alpha i + t} =  (\mathbf{I}_{N^\prime}\otimes {(\mathbf{a}^{(k)})}^\dagger) ({\mathbf{E}}_i\otimes \mathbf{E}_t)(\mathbf{I}_{N^\prime}\otimes \mathbf{a}^{(j)})) $, and ${\mathbf{H}^\prime}_{\alpha i + t} = ((\mathbf{I}_{M^\prime}\otimes {(\mathbf{b}^{(k)})}^\dagger)\hat{\mathbf{H}}_{\alpha i + t}(\mathbf{I}_{M^\prime}\otimes \mathbf{b}^{(j)})) $.

Observe that ${\mathbf{A}}_{\alpha i + t}$ only survives when $t = a^{(j)}$; then, using the mixed-product property 
\begin{equation}
\begin{aligned}
    {\mathbf{A}}_{\alpha i + t} &= \delta_{t,a^{(j)}}{\mathbf{E}}_i\otimes {(\mathbf{a}^{(k)})}^\dagger\mathbf{a}^{(j)}\\
    & = \delta_{a^{(k)},a^{(j)}}\delta_{t,a^{(j)}}{\mathbf{E}}_i.
\end{aligned}
\label{equ:ffaewf}
\end{equation} 
Substituting \eqref{equ:ffaewf} into the sum in \eqref{equ:summ}, we obtain
\begin{equation}
\begin{aligned}
{(\mathbf{Q}^{(k)})}^\dagger\mathbf{H}_{\text{TF}}\mathbf{Q}^{(j)}&=
    \sum_{i=0}^{N^\prime-1}\sum_{t=0}^{\alpha-1}\delta_{a^{(k)},a^{(j)}}\delta_{t,a^{(j)}}{\mathbf{E}}_i \otimes  {\mathbf{H}^\prime}_{\alpha i + t} \\
    &=\delta_{a^{(k)},a^{(j)}}\sum_{i=0}^{N^\prime-1}{\mathbf{E}}_i \otimes  {\mathbf{H}^\prime}_{\alpha i + a^{(k)}},
\end{aligned}
\end{equation}
which vanishes when $a^{(k)} \neq a^{(j)}$, yielding {\textbf{Lemma 3}}.

\textit{Remark 1}: This result implies that there is no mutual interference between users assigned to different time indices. Consequently, the level of \ac{MUI} can be effectively controlled by appropriately selecting the allocation parameters.

\begin{figure*}[t]
\centering
\includegraphics[scale = .99]{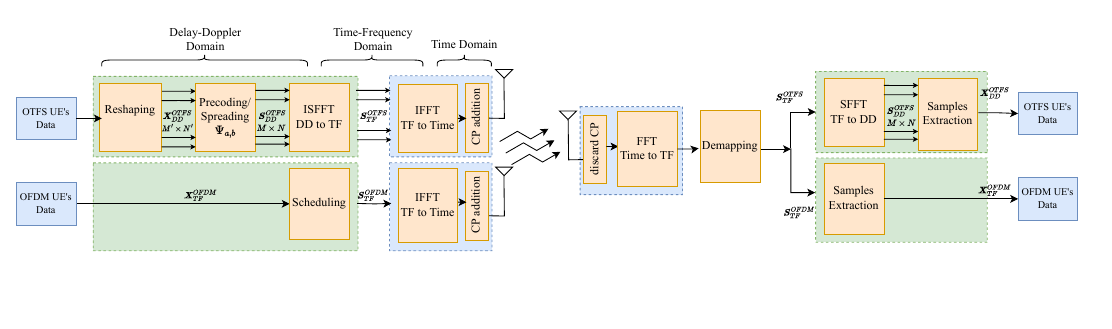}
\caption{Illustration of the UL system involving UEs of both waveforms according to the proposed scheme.}
\label{fig:block_diagram}
\end{figure*}

Figure~\ref{fig:block_diagram} shows the process of handling the samples of each user in an UL scenario. Transmission is performed according to \eqref{equ:mld} for OFDM users, and according to \eqref{equ:good_day} for OTFS users, as previously described. At the receiver, user separation is performed using the Hermitian transpose of the corresponding allocation matrix. For OFDM users, this operation yields the received data symbols directly. For OTFS users, the reception process is detailed in Section~\ref{sec:BER_analysis}.

\section{Performance Analysis}
In this section, we analyze the \ac{MUI} of the proposed OFDM–OTFS coexistence scheme and evaluate the performance of this scheme in terms of achievable rates, \ac{BER}, and sensing capabilities.
\subsection{Multi User Interference and Channel Conditions}
We examine the interference from the $j$-th UE to an arbitrary $k$-th UE that shares the same time index. At the \ac{BS}, the interference power \cite{tse2005fundamentals} observed over the $k$-th UE’s resources from the $j$-th UE is given by the squared Frobenius norm of $ {(\mathbf{Q}^{(k)})}^\dagger\mathbf{H}_{\text{TF}}\mathbf{Q}^{(j)} $, as 
\begin{equation}
    \text{T}_{k,j} \overset{\triangle}{=} \text{Tr} \big[ {\big( {(\mathbf{Q}^{(k)})}^\dagger\mathbf{H}_{\text{TF}}\mathbf{Q}^{(j)}\big)}^\dagger \big( {(\mathbf{Q}^{(k)})}^\dagger\mathbf{H}_{\text{TF}}\mathbf{Q}^{(j)}\big) \big],
\end{equation} 
where $\mathbf{H}_{\text{TF}}$ denotes the \ac{UL} channel for the $j$-th UE. As shown in \textbf{Lemma 3}, if users occupy different time indices, $\text{T}_{k,j}$ is zero. However, when both users share the same time resource, then whether or not $\text{T}_{k,j}$ is zero depends on channel conditions. 
We proceed by expanding the expression as follows:
\begin{equation}
    \begin{aligned}
        \text{T}_{k,j} &= \text{Tr}\bigg(\sum_{i=0}^{N^\prime-1}{\mathbf{E}}_i \otimes  {{\mathbf{H}^\prime}^\dagger_{\alpha i + a^{(k)}}}\sum_{i=0}^{N^\prime-1}{\mathbf{E}}_i \otimes  {\mathbf{H}^\prime}_{\alpha i + a^{(k)}}\bigg)\\
        &\overset{a}{=}\text{Tr}\bigg(\sum_{n=0}^{N^\prime-1}\sum_{m=0}^{N^\prime-1}{\mathbf{E}}_m{\mathbf{E}}_n \otimes  {{\mathbf{H}^\prime}^\dagger_{\alpha m + a^{(k)}}}{\mathbf{H}^\prime}_{\alpha n + a^{(k)}}\bigg)\\
        &\overset{b}{=}\text{Tr}\bigg(\sum_{n=0}^{N^\prime-1}{\mathbf{E}}_n \otimes  {{\mathbf{H}^\prime}^\dagger_{\alpha n + a^{(k)}}}{\mathbf{H}^\prime}_{\alpha n + a^{(k)}}\bigg)\\
        &\overset{c}{=}\sum_{n=0}^{N^\prime-1}\text{Tr}\left({{\mathbf{H}^\prime}^\dagger_{\alpha n + a^{(k)}}}{\mathbf{H}^\prime}_{\alpha n + a^{(k)}}\right)\\
        &\overset{d}{=}\sum_{n=0}^{N^\prime-1}\text{Tr}\big((\mathbf{I}_{M^\prime}\otimes \mathbf{b}^{(j)})(\mathbf{I}_{M^\prime}\otimes {(\mathbf{b}^{(j)})}^\dagger)\hat{\mathbf{H}}^\dagger_{\alpha n + a^{(k)}}\\
        & \quad\quad\times (\mathbf{I}_{M^\prime}\otimes \mathbf{b}^{(k)})(\mathbf{I}_{M^\prime}\otimes {(\mathbf{b}^{(k)})}^\dagger)\hat{\mathbf{H}}_{\alpha n + a^{(k)}}\big).\\
    \end{aligned}
     \label{equ:trace}
\end{equation}
The steps in~\eqref{equ:trace} can be explained as follows: (a) is due to the mixed-product property and the combination of the two summations; (b) is because basis matrices are orthogonal and are idempotent; (c) is due to the linearity of the trace operation and trace property of the Kronecker product; and (d) is due to the cyclic property of the trace \cite{van2000ubiquitous}.

OTFS UEs sharing the same time resources are the primary source of interference. To demonstrate this, we conduct a quantitative analysis of the interference. In the case of a general doubly dispersive channel, the trace can be derived using \eqref{equ:trace}, yielding a closed-form expression. To facilitate this derivation, \eqref{equ:trace} is first reformulated as follows:
\begin{equation}\small
    \begin{aligned}
        &\text{T}_{k,j} = \text{Tr}\bigg(\sum_{i=0}^{N^\prime-1}{\mathbf{E}}_i \otimes  {{\mathbf{H}^\prime}^\dagger_{\alpha i + a^{(k)}}}\sum_{i=0}^{N^\prime-1}{\mathbf{E}}_i \otimes  {\mathbf{H}^\prime}_{\alpha i + a^{(k)}}\bigg)\\
         &\quad=\sum_{n=0}^{N^\prime-1}\text{Tr}\bigg((\mathbf{I}_{M^\prime}\otimes \mathbf{B}^{(j)})\hat{\mathbf{H}}^\dagger_{\alpha n + a^{(k)}}(\mathbf{I}_{M^\prime}\otimes \mathbf{B}^{(k)})\hat{\mathbf{H}}_{\alpha n + a^{(k)}}\bigg)\\
        &\quad=\sum_{n=0}^{N^\prime-1}\text{Tr}\bigg({\big(\hat{\mathbf{H}}_{\alpha n + a^{(k)}}(\mathbf{I}_{M^\prime}\otimes \mathbf{B}^{(j)})\big)}^\dagger(\mathbf{I}_{M^\prime}\otimes \mathbf{B}^{(k)})\hat{\mathbf{H}}_{\alpha n + a^{(k)}}\bigg),\\
    \end{aligned}
\end{equation}
where $\mathbf{B}^{(i)} = \mathbf{E}_{b^{(i)}} $  is the $b^{(i)}$-th basis matrix of size $\beta \times \beta$, which was obtained through the use of the mixed product property on (d) from \eqref{equ:trace}. The matrix multiplication with $\mathbf{B}^{(i)}$ on the left selects the $b^{(k)}$-th column, while multiplication on the right selects the $b^{(j)}$-th row.
By definition, the trace can be expressed as follows:
\begin{equation}\small
    \begin{aligned}
        \text{T}_{k,j} &= \sum_{n=0}^{N^\prime-1}\text{Tr}({[\hat{\mathbf{H}}_{\alpha n + a^{(k)}}(\mathbf{I}_{M^\prime}\otimes \mathbf{B}^{(j)})]}^\dagger(\mathbf{I}_{M^\prime}\otimes \mathbf{B}^{(k)}))\hat{\mathbf{H}}_{\alpha n + a^{(k)}}))\\
         &=\sum_{n=0}^{N^\prime-1}\sum_{m_1=0}^{M^\prime-1}\sum_{m_2=0}^{M^\prime-1}{\left|\hat{\mathbf{H}}_{\alpha n + a^{(k)}}(m_1\beta+ b^{(k)},m_2\beta + b^{(j)}) \right|}^2.
    \label{equ:gentrace}
  \end{aligned}
\end{equation}

Under the assumption that the delays along different paths are distinct and are independent of both $a^{(k)}$ and $n$, a closed-form expression for $\text{T}_{k,j}$ can be derived. In particular, we obtain
\begin{equation}
    \begin{aligned}
        \sum_{m_1=0}^{M^\prime-1}&\sum_{m_2=0}^{M^\prime-1}{\left|\hat{\mathbf{H}}_{\alpha n + a^{(k)}}(m_1\beta+ b^{(k)},m_2\beta + b^{(j)}) \right|}^2 \\
        &=\sum_{l_i \equiv b \: (\mathrm{mod}  \: \beta) } M'{|h^{(j)}_i|}^2 ,
  \end{aligned}
\end{equation} 
where $b = b^{(k)}-b^{(j)}$. Then, 
\begin{equation}
    \begin{aligned}
        \text{T}_{k,j}
        =\sum_{l_i \equiv b \: (\mathrm{mod}  \: \beta)}N'M'{|h^{(j)}_i|}^2 .
        \label{equ:trace_15_10}
  \end{aligned}
\end{equation}

\textit{Remark 2}:
Here it is assumed that users $k$ and $j$ overlap in time, since otherwise the trace would be zero as shown in {\textbf{Lemma 3}}. 

\textit{Remark 3}:
Under the assumption of a channel model with an exponentially decaying \ac{PDP} \cite{goldsmith2005wireless}, the interference level $\text{T}_{k,j}$ can be mitigated by increasing the frequency separation $b = b^{(k)}-b^{(j)}$ between the two users. This is because larger $b$ values correspond to larger delay indices $l_i$ that satisfy the congruence condition $l_i \equiv b  \:     (\mathrm{mod}  \: \beta) $, which, under the assumed PDP, generally result in exponentially smaller channel gain magnitudes.

To proceed with the performance evaluation, we derive an expression for the \ac{SINR}, which directly impacts the achievable rate of each user. Therefore, it is important to have an expression for the desired signal strength in addition to the interference strength. In particular, the signal strength of the $k$-th UE is given by $~\text{T}_{k,k}$. In this case, $b=0$ and $l_i = v\cdot \beta$ for any integer $v$. Substituting this condition into the trace expression yields
 \begin{equation}
        \text{T}_{k,k}
        =\sum_{l_i = v\cdot \beta}N'M'{|h^{(k)}_i|}^2 
        =N'M'\left({|h^{(k)}_0|}^2+{|h^{(k)}_\beta|}^2+\cdots\right).
   \label{equ:SS}
\end{equation}

\textit{Special case:} For a static channel (i.e., $k_i=0$), only delay-spread is present and the channel is no longer doubly dispersive. Under this condition, the diagonal blocks given in~\eqref{equ:Ht} reduce to $M \times M$ circulant matrices. These circulant matrices can be diagonalized using the \ac{DFT} matrix as follows:
\begin{equation}
    \begin{aligned}
        \mathbf{H}_{n} =\mathbf{F}_M^\dagger\mathbf{D}_n\mathbf{F}_M,
    \end{aligned}
\end{equation}
where $\mathbf{D}_n$ is an $M\times M$ diagonal matrix. Then, from \eqref{equ:momo}, the \Ac{TF} domain blocks become diagonal as 
\begin{equation}
    \begin{aligned}
        \hat{\mathbf{H}}_{n} &=\mathbf{F}_M\mathbf{H}_{n}\mathbf{F}_M^\dagger\\
        &=\mathbf{D}_n.\\
    \end{aligned}
     \label{equ:dddd}
\end{equation}
Substituting \eqref{equ:dddd} in \eqref{equ:trace}, we find
\begin{equation}
    \begin{aligned}
        \text{T}_{k,j} =& 
        \sum_{n=0}^{N^\prime-1}\text{Tr}((\mathbf{I}_{M^\prime}\otimes \mathbf{b}^{(j)}{(\mathbf{b}^{(j)})}^\dagger){\mathbf{D}}^\dagger_{\alpha n + a^{(k)}}\\
        & \times (\mathbf{I}_{M^\prime}\otimes \mathbf{b}^{(k)}{(\mathbf{b}^{(k)})}^\dagger){\mathbf{D}}_{\alpha n + a^{(k)}}).
    \end{aligned}
\end{equation}
This result corresponds to the trace of a product of diagonal matrices, which can be computed by summing the element-wise products of their diagonal entries. Since the two users occupy different frequency indices, the nonzero diagonal entries of $\mathbf{I}_{M^\prime}\otimes \mathbf{b}^{(k)}{(\mathbf{b}^{(k)})}^\dagger$ coincide with the zeros in $\mathbf{I}_{M^\prime}\otimes \mathbf{b}^{(j)}{(\mathbf{b}^{(j)})}^\dagger$, and vice versa. Thus, the sum of their element-wise product is zero. This gives zero trace, which confirms the intuition that static users cause no leakage. Moreover, when users are assigned different time indices, the trace is zero regardless of the channel conditions, as established by \textbf{Lemma 3}.

\textbf{Corollary 1}:
In the case where the $j$-th UE is a \ac{LM} user with a time-invariant channel, it causes no interference to an arbitrary $k$-th UE. That is
\begin{equation}
    \text{T}_{k,j} = 0.
\end{equation}

The results presented in this subsection form the foundation for evaluating the performance of the proposed scheme. In the next subsection, these results will be contextualized, systematically analyzed, and interpreted to provide deeper insights into their implications.

\subsection{Achievable Rates Analysis}

In the considered coexistence paradigm, users experiencing doubly dispersive channels are assigned OTFS modulation, while users with time-invariant channels are served using OFDM. In a \ac{DL} scenario, only the OTFS UEs experience \ac{MUI} depending on the chosen allocation parameters as discussed in the previous section. In this section, we focus on an \ac{UL} scenario involving users with varying mobility profiles. Each user experiences a distinct channel, and the aggregate received signal at the \ac{BS} is expressed in \eqref{equ:in_out_ul}. 

The received signal corresponding to the $k$-th user is found by multiplying the received signal vector with the Hermitian of the corresponding allocation matrix $\mathbf{Q}^{(k)}$. This gives
\begin{equation}
\begin{aligned} 
 \mathbf{y}^{(k)} &={(\mathbf{Q}^{(k)})}^\dagger\left(\sum_{j=1}^{U}\mathbf{H}^{(j)}_{\text{TF}}\mathbf{Q}^{(j)}\mathbf{x}^{(j)} + \mathbf{w}\right)\\
 &=\sum_{j=1}^{U}{(\mathbf{Q}^{(k)})}^\dagger\mathbf{H}^{(j)}_{\text{TF}}\mathbf{Q}^{(j)}\mathbf{x}^{(j)} + {(\mathbf{Q}^{(k)})}^\dagger\mathbf{w}\\
 &={(\mathbf{Q}^{(k)})}^\dagger\mathbf{H}^{(k)}_{\text{TF}}\mathbf{Q}^{(k)}\mathbf{x}^{(k)}\\
 &+\underbrace{\sum_{j=1,j \neq k}^{U}{(\mathbf{Q}^{(k)})}^\dagger\mathbf{H}^{(j)}_{\text{TF}}\mathbf{Q}^{(j)}\mathbf{x}^{(j)} +\mathbf{w}^\prime}_{\text{interference}},
 \label{equ:UL_rx_k}
\end{aligned}
\end{equation}

where $\mathbf{w}'={(\mathbf{Q}^{(k)})}^\dagger\mathbf{w}$ denotes the noise vector projected onto the $k$-th user’s subspace.

This creates parallel user-specific channels in the \ac{TF} domain, where each UE’s signal passes through an effective channel given in the third line of \eqref{equ:UL_rx_k}.  The last line in \eqref{equ:UL_rx_k} represents the interference from all users on the $k$-th user. As shown in \textbf{Corollary 1}, the only non-zero interference arises from OTFS UEs sharing the same time index, and its magnitude depends on the specific allocation parameters. Assuming the original noise vector $\mathbf{w} \sim \mathcal{CN}(\mathbf{0}, \sigma^2 \mathbf{I})$ is \ac{iid} complex Gaussian, the projected noise $\mathbf{w}'$ follows a complex Gaussian distribution with covariance matrix $\mathbb{E}[\mathbf{w}^\prime \mathbf{w}^{\prime \dagger}] = \sigma^2 {(\mathbf{Q}^{(k)})}^\dagger \mathbf{Q}^{(k)}= \sigma^2\mathbf{I}$ and remains i.i.d. \ac{AWGN} with the same variance per dimension. For simplicity, let
\begin{equation}
 \mathbf{\Xi}_{(k)}^{(j)} \triangleq{(\mathbf{Q}^{(k)})}^\dagger\mathbf{H}^{(j)}_{\text{TF}}\mathbf{Q}^{(j)}.
\end{equation} 
Then, the expression for the \ac{SINR} for the $k$-th user can be given as
\begin{equation}
     \text{SINR}^{(k)} = 
    \frac{E_0\text{Tr}({\mathbf{\Xi}_{(k)}^{(k)}}^\dagger\mathbf{\Xi}_{(k)}^{(k)})}{N_0 +{\text{Tr}\left(\sum_{j=1,j \neq k}^{U-1}E_0{\mathbf{\Xi}_{(k)}^{(j)}}^\dagger\mathbf{\Xi}_{(k)}^{(j)}\right)}},
    \label{eq:ul_sinr_def}
\end{equation}
where $E_0$ is the average symbol energy, and $ N_0$ is the noise power spectral density. Next, this SINR expression is simplified, and a closed-form upper bound on the achievable rate is derived.

\subsubsection{Upper Bounds on Achievable Rates for the Proposed Scheme}
The achievable rate for the $k$-th user is expressed as \cite{tse2005fundamentals}
\begin{equation}
    R^{(k)} = \frac{1}{M'N'}\log_2\det\left( \mathbf{I}_{M^\prime N^\prime} +\frac{E_0}{N_0 +\epsilon_k} {\mathbf{\Xi}_{(k)}^{(k)}}^{\dagger}{\mathbf{\Xi}}_{(k)}^{(k)} \right),
    \label{equ:rate_definition}
\end{equation}
where 
\begin{equation}
     \epsilon_k = \frac{1}{M'N'}
    {\text{Tr}\left(\sum_{j=1,j \neq k}^{U-1}E_0{\mathbf{\Xi}_{(k)}^{(j)}}^\dagger\mathbf{\Xi}_{(k)}^{(j)}\right)}.
\end{equation}
The matrix ${\mathbf{\Xi}_{(k)}^{(k)}}^{\dagger}{\mathbf{\Xi}}_{(k)}^{(k)}$ is positive semi-definite. Hence, its eigenvalues are real and non-negative, and the matrix $\mathbf{I}_{M^\prime N^\prime} +\frac{E_0}{N_0 +\epsilon_k} {\mathbf{\Xi}_{(k)}^{(k)}}^{\dagger}{\mathbf{\Xi}}_{(k)}^{(k)}$ is positive-definite. Let the eigen-values of ${\mathbf{\Xi}_{(k)}^{(k)}}^{\dagger}{\mathbf{\Xi}}_{(k)}^{(k)}$ be $\lambda^{(k)}_1,\lambda^{(k)}_2,...,\lambda^{(k)}_{M^\prime N^\prime}$. Using eigen-value decomposition (EVD), \eqref{equ:rate_definition} can be rewritten as follows:
\begin{equation}
    R^{(k)} = \frac{1}{M'N'}\sum_{i=1}^{M^\prime N^\prime}\log_2\left( 1 + \frac{E_0}{N_0 +\epsilon_k} \lambda^{(k)}_i \right).
\end{equation}

At this point, an upper bound for $R^{(k)}$ can be derived using Jensen’s inequality. The inequality states that for a concave function $\phi(x)$ and non-negative real numbers $x_1, x_2, \dots, x_n$, the following holds

\begin{equation}
     \frac{1}{n}\sum_{i=1}^{n}\phi\left( x_i \right) \leq \phi\left( \frac{1}{n}\sum_{i=1}^{n}x_i \right).
\end{equation}
It is straight forward to show that the function $\log(1+x)$ is concave (since its derivative $\frac{1}{1+x}$ is strictly monotonically decreasing over the interval of interest). Applying this inequality to the achievable rate expression, we obtain
\begin{equation}
\begin{aligned}
     R^{(k)} =\frac{1} {M'N'}&\sum_{i=1}^{M^\prime N^\prime}\log_2\left( 1 + \frac{E_0}{N_0 +\epsilon_k} \lambda^{(k)}_i \right) \\
     &\leq\log_2\left( 1 + \frac{E_0}{N_0 +\epsilon_k}\frac{\text{Tr}\left({\mathbf{\Xi}_{(k)}^{(k)}}^\dagger\mathbf{\Xi}_{(k)}^{(k)}\right)}{M^\prime N^\prime}  \right).
\end{aligned}
\end{equation}

Thus, an upper bound for $R^{(k)}$ is given by\footnote{Note that the trace is simply used to express the sum of the eigen values.}
\begin{equation}
     R^{(k)}_{\text{upper-bound}} =\log_2\left( 1 + \frac{E_0}{N_0 +\epsilon_k}\frac{\text{Tr}\left({\mathbf{\Xi}_{(k)}^{(k)}}^\dagger\mathbf{\Xi}_{(k)}^{(k)}\right)}{M^\prime N^\prime}  \right).
\end{equation}
To proceed, it is useful to define $\mathcal{U}_k$, the set of OTFS UEs who share the same time indices as the $k$-th user. From \eqref{equ:trace_15_10}, we have 
\begin{equation}
     \epsilon_k = 
    E_0\sum_{j\in \mathcal{U}_k}\sum_{l_i \equiv b \: (\mathrm{mod}  \: \beta)}{|h^{(j)}_i|}^2,
\end{equation}
where $b = b^{(k)}-b^{(j)}$. Using \eqref{equ:SS}, we can write
 \begin{equation}
\begin{aligned}
     R^{(k)}_{\text{upper-bound}} =\log_2\left( 1 + \frac{E_0}{N_0 +\epsilon_k}\sum_{l_i = v\cdot \beta} {|h^{(k)}_i|}^2  \right).\\
\end{aligned}
\end{equation}

As will be demonstrated through extensive simulations, the average rate performance achieved by the different users closely approaches this upper bound. From {\textbf{Lemma 3}}, it follows that $\epsilon_k$ accounts only for contributions from users occupying the same time indices. Additionally, as discussed in \textbf{Corollary 1}, $\epsilon_k$ does not include interference from OFDM UEs. Consequently, the users most affected are those sharing time indices with OTFS UEs.

For the sake of completeness, we can write the closed form expression for the resulting SINR. It is evident from the earlier derivation that \eqref{eq:ul_sinr_def} can be simplified to
\begin{equation}
     \text{SINR}^{(k)} = 
    \frac{E_0}{N_0 +\epsilon_k}\sum_{l_i = v\cdot \beta} {|h^{(k)}_i|}^2.
\end{equation}

\subsection{BER Performance Evaluation}\label{sec:BER_analysis}

The strength of \ac{OTFS} lies in its ability to fully exploit channel diversity in the \ac{DD} domain, an area where other waveforms, such as \ac{OFDM} and \ac{SC}, fall short. This property enables OTFS to achieve superior \ac{BER} performance. However, fully realizing this advantage requires that OTFS symbols span sufficiently wide bandwidths and long time durations, which is typically at odds with conventional coexistence strategies. In the proposed system, each user is allowed to access the maximum available bandwidth and time, thereby achieving the desired diversity and significantly improving \ac{BER} performance.
To demonstrate the BER gains obtained through spreading, we consider an alternative expression for the \ac{TF} channel\footnote{We will temporarily drop the UE index here, all the matrices correspond to the same UE.}. From \eqref{equ:time_domain_channel}, we can rewrite \eqref{equ:Ht} as 
\begin{equation}
    \mathbf{H}_{\text{T}} =\sum_{i=1}^{P} h_i\mathbf{H}'_i, 
\end{equation}
where $\mathbf{H}'_i = \text{diag}\left (\mathbf{H}_0,...,\mathbf{H}_{N-1}  \right)$ and $\mathbf{H}_{n} = \boldsymbol{\Pi}^{l_i}_{M}\boldsymbol{\Omega}^{k_i,n}_{M}.$

As before, $\mathbf{H}_{\text{T}}$ is block diagonal. Hence, by  \textbf{Lemma 1}, the corresponding \ac{TF} domain channel matrix $ \mathbf{H}_{\text{TF}} $ given by \eqref{equ:H_tf} is also block diagonal, with each block expressed as
\begin{equation}
    \hat{\mathbf{H}}_n = \mathbf{F}_M\mathbf{H}_{n} \mathbf{F}^\dagger_M.
\end{equation}
Accordingly, we can write \eqref{equ:Ht} as 
\begin{equation}
    \mathbf{H}_{\text{TF}} =\sum_{i=1}^{P} h_i\mathbf{H}_i, 
    \label{equ:H_equation_sum_of_paths}
\end{equation}
where $\mathbf{H}_i = (\mathbf{I}_N\otimes \mathbf{F}_M)\mathbf{H}'_{i} ( \mathbf{I}_N \otimes \mathbf{F}^\dagger_M).$

Although the \ac{TF} domain samples for an OTFS UE are not exactly the \ac{ISFFT} of the \ac{DD} domain symbols due to the phase term in \eqref{equ:spread_theorem}, the original symbols can still be effectively recovered through various methods.

Let $\mathbf{x}$ be the vector of transmitted symbols. From \eqref{eq_for_T}, the samples $\text{T}(n^\prime,m^\prime)$ can be expressed in vector form as
\begin{equation}
     \mathbf{t}= (\mathbf{F}'_N\otimes\mathbf{F}'_{M})\mathbf{x},
\end{equation}
where $\mathbf{F}'_{M}$ is a sub-matrix of $\mathbf{F}_{M}$ containing the first $M'$ columns, and  $\mathbf{F}'_{N}$ is the transpose of a sub-matrix of $\mathbf{F}^\dagger_{N}$ consisting of the first $N'$ rows.

One way to recover the \ac{DD} symbols would be to obtain the \ac{TF} samples corresponding to the user, then spread them back before taking the \ac{SFFT}. The symbols can be found in the first $M' \times N'$ sub-matrix.
This process can be modeled by 
\begin{equation}
    \mathbf{y} =\mathbf{Q}_\text{DD}(\mathbf{F}_N\otimes\mathbf{F}^\dagger_{M})\mathbf{Q}{\mathbf{Q}}^\dagger\mathbf{H}_{\text{TF}}\mathbf{Q}{\mathbf{Q}}^\dagger(\mathbf{F}'_N\otimes\mathbf{F}'_{M})\mathbf{x} + \Tilde{\mathbf{w}},
    \label{inout_symbol_relation}
\end{equation} 
where  $\mathbf{y}$ is the received symbol vector, $\Tilde{\mathbf{w}}$ represents the effect of the noise and the \ac{MUI} with combined power denoted by $\Tilde{N}_0$; and 
\begin{equation}
    \mathbf{Q}_\text{DD} =\mathbf{e}_\alpha\otimes\mathbf{I}_{N'}\otimes\mathbf{e}_\beta\otimes\mathbf{I}_{M'},
\end{equation} 
where $\mathbf{e}_k = [1,0,0,...,0] $ is a standard basis vector of size $k\times 1$. 

Combining this with \eqref{equ:H_equation_sum_of_paths}, we can write 
\begin{equation}
    \mathbf{y} =\sum_{i=1}^{P} h_i\Bar{\mathbf{H}}_i\mathbf{x} + \Tilde{\mathbf{w}} = \Phi(\mathbf{x})\mathbf{h}+\Tilde{\mathbf{w}},
\end{equation} 
where $\mathbf{h} = [h_1,h_2,...,h_P]$ is a $P\times 1$ vector and $\Phi(\mathbf{x})$ is an $N\times P$ matrix given by
\begin{equation}
    \Phi(\mathbf{x}) = [\Bar{\mathbf{H}}_i\mathbf{x}|...|\Bar{\mathbf{H}}_P\mathbf{x}],
\end{equation} 
and
\begin{equation}
    \Bar{\mathbf{H}}_i =\mathbf{Q}_\text{DD}(\mathbf{F}_N\otimes\mathbf{F}^\dagger_{M})\mathbf{Q}{\mathbf{Q}}^\dagger{\mathbf{H}_i}\mathbf{Q}{\mathbf{Q}}^\dagger(\mathbf{F}'_N\otimes\mathbf{F}'_{M}) . 
    \label{inout_symbol_relation}
\end{equation} 
The \ac{PEP} can then be expressed as follows \cite{biglieri2019error}: 

\begin{equation}
\begin{aligned}
\operatorname{P}(\mathbf{x}\longrightarrow\mathbf{x}') = \operatorname{E}\left[Q\left(\sqrt{\frac{{||\Phi(\mathbf{x}'-\mathbf{x})\mathbf{h}||}^2}{2\Tilde{N}_0}} \right)\right],
\end{aligned}
\end{equation}
where the averaging is over channel realizations.

The term ${||\Phi(\mathbf{x}'-\mathbf{x})\mathbf{h}||}^2$ can be expressed as ${\mathbf{h}^\dagger{\Phi(\mathbf{x}'-\mathbf{x})}^\dagger}{\Phi(\mathbf{x}'-\mathbf{x})\mathbf{h}}$. 
This Hermitian matrix can be diagonalized by a unitary matrix $\mathbf{\Delta}$, i.e., 
${\Phi(\mathbf{x}'-\mathbf{x})}^\dagger{\Phi(\mathbf{x}'-\mathbf{x})} = \mathbf{\Delta}^{\dagger}\mathbf{\Lambda}\mathbf{\Delta}$, where $\mathbf{\Lambda} = \text{diag}\{\lambda^2_1,\lambda^2_2,...,\lambda^2_P \}$ with $\lambda_1,\lambda_2,...,\lambda_P$ being the singular values of $\Phi(\mathbf{x}'-\mathbf{x})$.

Define $\mathbf{h}' = \mathbf{\Delta}\mathbf{h}$, and let $r$ denote the rank of $\Phi(\mathbf{x}'-\mathbf{x})$. For simplicity of the presentation, assume $\Tilde{\mathbf{w}}$ is normally distributed with zero mean.

Then, we can express the \ac{PEP} as 
\begin{equation}
\begin{aligned}
\operatorname{P}(\mathbf{x}\longrightarrow\mathbf{x}') = \operatorname{E}\left[Q\left(\sqrt{\frac{\sum_{i=1}^{r}\lambda^2_i{|{h}'_i|}^2}{2\Tilde{N}_0}} \right)\right].
\end{aligned}
\end{equation}

To better understand the performance of the proposed spreading, assume that \ac{BPSK} modulation is employed, with equiprobable symbols from the modulation alphabet. A lower bound for the \ac{BER} can be derived by identifying pairs of symbol vectors $\mathbf{x}'$ and $\mathbf{x}$ for which $\Phi(\mathbf{x}' - \mathbf{x})$ is a rank-one matrix. Denoting the set of such pairs of vectors by $\mathcal{A}$ and the size of this set by $\aleph$, the lower bound can be found from\cite{surabhi2019diversity} as

\begin{equation}
\begin{aligned}
 \textbf{BER} \geq & \frac{1}{2^{MN}}\sum_{(\mathbf{x},\mathbf{x}') \in \mathcal{A}}\operatorname{P}(\mathbf{x}\longrightarrow\mathbf{x}')\\
 \label{BER_lower_bound_1}
\end{aligned}
\end{equation}
In this case, there would be a single nonzero eigen value, corresponding to each pair $(\mathbf{x},\mathbf{x}')$. Let $\lambda$ be the largest eigen value for all pairs in $\mathcal{A}$. Then,
\begin{equation}
\begin{aligned}
\operatorname{P}(\mathbf{x}\longrightarrow\mathbf{x}') & = \operatorname{E}\left[Q\left(\sqrt{\frac{\sum_{i=1}^{r}\lambda^2_i{|{h}'_i|}^2}{2\Tilde{N}_0}} \right)\right]\\
&\geq \operatorname{E}\left[Q\left(\sqrt{\frac{\lambda^2{|{h'_1}|}^2}{2\Tilde{N}_0}} \right)\right].
\end{aligned}
\end{equation}
Assuming that ${h}_1 \sim \mathcal{C}\mathcal{N}(0, 1/P)$, then ${h}'_1 \sim \mathcal{C}\mathcal{N}(0, 1/P) $, since $\mathbf{\Delta}$ is unitary. Therefore,
\begin{equation}
 \operatorname{E}\left[Q\left(\sqrt{\frac{\lambda^2{|{h'_1}|}^2}{2\Tilde{N}_0}} \right)\right]=\frac{1}{2}\left( 1-\sqrt{\frac{\lambda}{\lambda+P\Tilde{N}_0}} \right).
 \label{equ:dasdasd}
\end{equation}
Substituting \eqref{equ:dasdasd} into \eqref{BER_lower_bound_1}, we find
\begin{equation}
 \textbf{BER} \geq \frac{\aleph}{2^{MN}}\frac{1}{2}\left( 1-\sqrt{\frac{\lambda}{\lambda+P\Tilde{N}_0}} \right),
 \label{equ:BER_Lower_bound_pre_finally}
\end{equation}
which, at high \text{SNR}, simplifies to 
\begin{equation}
\begin{aligned}
 \textbf{BER} \geq \frac{\aleph}{2^{MN}}\frac{\Tilde{N}_0P}{4 \lambda}.
 \label{equ:BER_Lower_bound_finally}
\end{aligned}
\end{equation}
This bound is influenced by various factors, including interference and the structure of the matrices involved. However, the key insight lies in its dependence on the values of $M$ and $N$, which exhibits a consistent trend across all UEs. 
From \eqref{equ:BER_Lower_bound_pre_finally} and \eqref{equ:BER_Lower_bound_finally}, the factor $\frac{1}{2^{MN}}$ highlights a substantial potential gain of $3\alpha\beta~\mathrm{dB}$ compared to unspread schemes.
 
Naturally, the bound will differ for higher-order modulation schemes; however, the trend remains consistent. For higher values of $M$ and $N$, the bound is smaller and the \ac{BER} performance is enhanced.

\begin{figure}[t]
  \centering
     \subfigure[OTFS scheduling in \cite{wu2022co}.]{\includegraphics[scale=0.33]{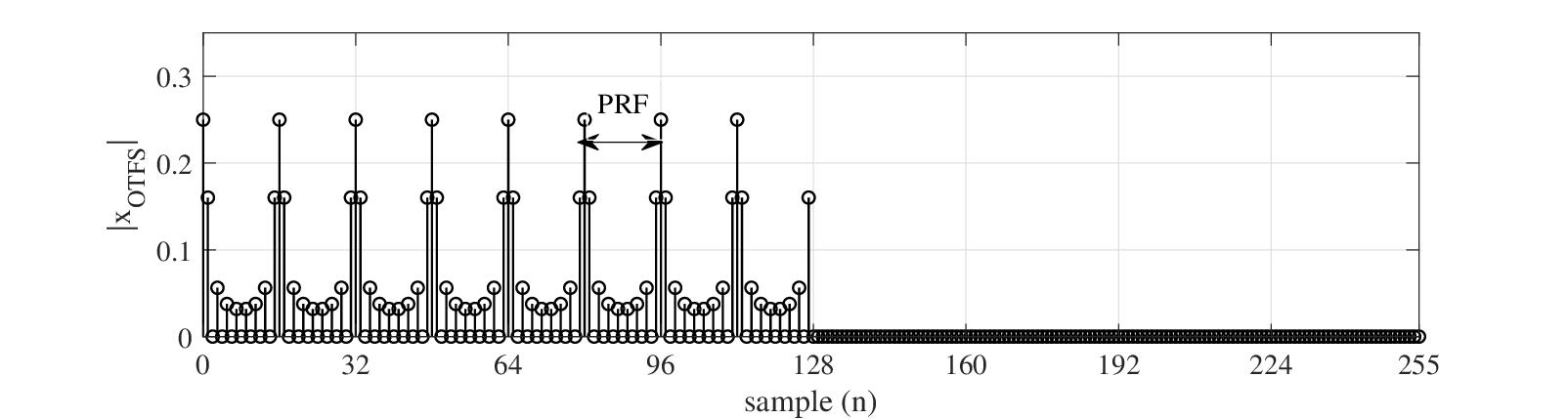}}
     \subfigure[Proposed OTFS scheduling ($\alpha = \beta = 2$).]{\includegraphics[scale=0.33]{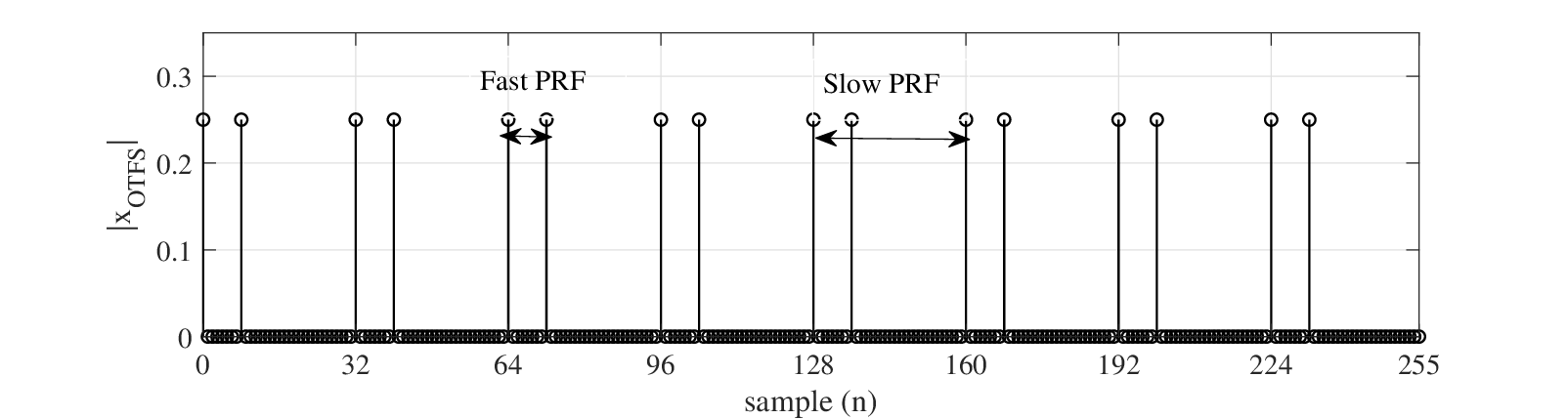}} 
  \caption{OTFS carrier time signal sampled at $f_s = \frac{1}{B}$ for duration $T$.}   
      \label{fig:sensing_illustration}
 \end{figure}

\subsection{Sensing Performance Evaluation}

This subsection evaluates and compares the sensing capabilities of two OTFS systems: one employing conventional scheduling and the other utilizing the proposed spreading scheme. The comparison is conducted with respect to key radar metrics, including range resolution $\Delta R$, velocity resolution $\Delta V$, maximum unambiguous range $R_{\max}$, and maximum unambiguous velocity $V_{\max}$. 

\subsubsection{Range Resolution (\(\Delta R\))}
The range resolution is defined by $\Delta R = \frac{c}{2B}$
where \(c\) is the speed of light (\(c = 3 \times 10^8 \, \text{m/s}\)). In conventional coexistence systems, the effective bandwidth is reduced to \(B/\beta\), resulting in \cite{liu2020super}
\begin{equation}
\Delta R_{\text{conv}} = \frac{c}{2 \cdot (B/\beta)} = \beta \cdot \Delta R_{\text{prop}}.
\end{equation}
In contrast, the proposed system leverages the full bandwidth \(B\), providing \(\beta\)-times better range resolution.

\subsubsection{Velocity Resolution (\(\Delta V\))} 
The velocity resolution is given by $\Delta V = \frac{\lambda_c}{2T}$, where $\lambda_c = \frac{c}{f_c}$ is the carrier wavelength. For the conventional system, the effective time duration is reduced to \(T/\alpha\), yielding
\begin{equation}
\Delta V_{\text{conv}} = \frac{\lambda_c}{2 \cdot (T/\alpha)} = \alpha \cdot \Delta V_{\text{prop}}.
\end{equation}
The proposed system uses the full time duration \(T\), offering \(\alpha\)-times better velocity resolution.

\subsubsection{Maximum Unambiguous Range (\(R_{\text{max}}\))}  
The maximum unambiguous range is defined as $R_{\text{max}} = \frac{c}{2 f_{\text{PR}}}$, where $f_{\text{PR}}$ denotes the pulse repetition frequency (PRF).
Compared to the conventional system, shown in Fig. \ref{fig:sensing_illustration}(a), the pulse repetition becomes faster, increasing the PRF to \(f_{\text{PR},f} = \alpha\cdot f_{\text{PR}}\) in the proposed scheme, as depicted in Fig. \ref{fig:sensing_illustration}(b), leading to
\begin{equation}
R_{\text{max, prop}} = R_{\text{max, conv}}/\alpha .
\end{equation}
This indicates that the unambiguous range is reduced by a factor of $\alpha$.

\subsubsection{Maximum Unambiguous Velocity (\(V_{\text{max}}\))}  
The maximum unambiguous velocity is given by $V_{\text{max}} = \frac{f_{\text{PR}} \lambda_c}{4}$. In the proposed system, the PRF is reduced, depicted by the slow PRF in Fig. \ref{fig:sensing_illustration}(b), to \(f_{\text{PR},s} = f_{\text{PR}}/\beta\), leading to
\begin{equation}
V_{\text{max, prop}} =  V_{\text{max, conv}}/\beta.
\end{equation}
Hence, the proposed system also sacrifices maximum unambiguous velocity, reducing it by a factor of $\beta$.

\subsubsection{Conclusion and implications for ISAC}
The proposed system enhances both range and velocity resolutions by factors of $\beta$ and $\alpha$, respectively. This improvement is achieved by utilizing the full bandwidth \(B\) and time duration \(T\). However, this gain comes at the expense of reduced PRF, which in turn decreases the maximum unambiguous range and velocity.  The proposed system represents a deliberate trade-off that prioritizes sensing resolution over ambiguity metrics. This trade-off makes the proposed scheme more suitable for ISAC applications where precise sensing is of higher importance than extended unambiguous range or velocity.

The sensing evaluation presented here focuses on the fundamental resolution limits enabled by the proposed resource allocation and waveform structure. As such, the results isolate and emphasize the fundamental trade-off between resolution and unambiguous range/velocity due to spreading. Practical challenges (e.g., channel estimation in the presence of interference, pilot contamination) are beyond this paper’s scope. Incorporating a complete sensing chain remains a subject for future work.

 \begin{table} [t]
\begin{center}
\caption{Simulation Parameters} 
\label{tab:sim-parm}
\begin{tabular}{l|c|c}
\hline \textbf{Description} & \textbf{Parameters} & \textbf{Value}  \\
\hline  Delay, Doppler & $N, M$ & (32,32) \\
\hline  Spreading factors ``prop 1'' & $(\alpha, \beta)$ & $(4,1)$ \\
\hline  Spreading factors ``prop 2'' & $(\alpha, \beta)$ & $(1,4)$ \\
\hline  Spreading factors ``prop 3'' & $(\alpha, \beta)$ & $(2,2)$ \\
\hline Benchmark\cite{wu2022co} ``sch 1'' & TDMA &   \\
\hline Benchmark\cite{wu2022co} ``sch 2'' & FDMA &   \\
\hline Benchmark\cite{wu2022co} ``sch 3'' & TDMA+FDMA &   \\
\hline Number of UEs & $U$ & $4$  \\
\hline
\end{tabular}
\end{center}
\end{table}

\section{Simulation Results}

In this section, the performance of the proposed coexistence scheme is evaluated through simulation. For the achievable rate and BER simulations, we consider a scenario in which an \ac{OTFS} user coexists with multiple \ac{OFDM} UEs. The key system parameters employed in the simulations are summarized in Table~\ref{tab:sim-parm}.
The parameters $M$, $N$, and the spreading factors are selected for simulation convenience and can be appropriately scaled for practical deployments. Throughout the evaluation, we focus on three users out of the $U$ total users: the ``OTFS User''; an arbitrary OFDM user (``Arbitrary User'') assigned to a frequency resource that is nonadjacent to the OTFS user, and the worst-case OFDM user (``Worst User'') assigned to a frequency resource directly adjacent to the OTFS user.

\begin{figure}[t]
\centering
\includegraphics[scale = .25]{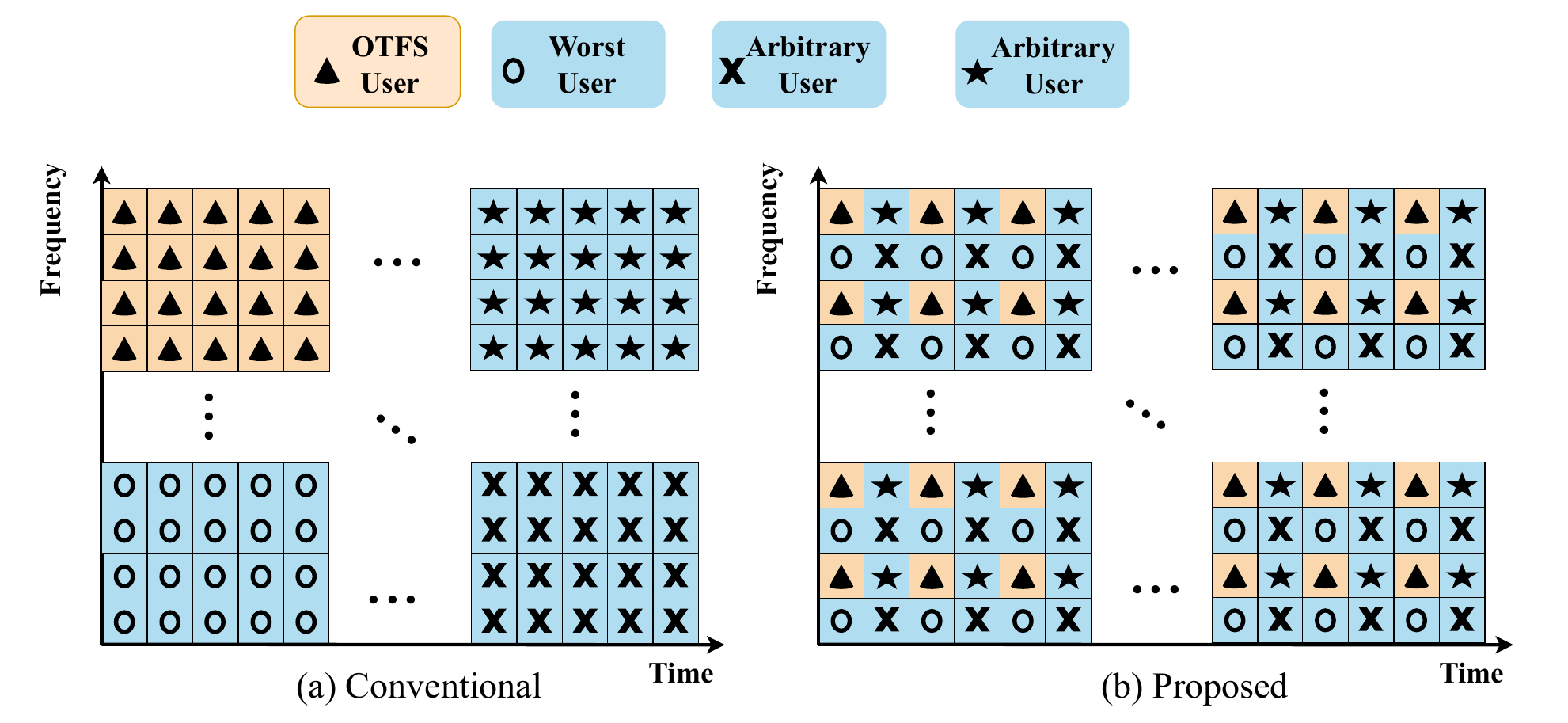}
\caption{The allocation of users for the considered scenario.}
\label{fig:simulation_scenario}
\end{figure}

 An example of the allocation of all four users is shown in Fig.~\ref{fig:simulation_scenario}, for the orthogonal schemes (in this example, $\alpha = \beta = 2$). The OFDM users are shown in blue, and either one of the two UEs labeled ``Arbitrary User'' can be chosen to represent the arbitrary OFDM user. We examine different choices of the spreading parameters\footnote{Different choices of the spreading parameters can represent different scenarios involving an OTFS user coexisting with a number of OFDM users. In scenarios involving more than a single OTFS user, $\alpha$ and $\beta$ alone are not enough, as the allocation of OTFS users is also important. This allocation has to be optimized based on all users' requirements, by defining specific quality-of-service metrics.}. We consider three options. First, ``prop 1'' and its unspread conventional version ``sch 1'' are \ac{TDMA}-based allocations ($\beta =1$). Secondly, ``prop 2'' and its unspread counterpart ``sch 2'' are \ac{FDMA}-based allocations ($\alpha =1$). Lastly, ``prop 3'' and ``sch 3'' are a combination of the previous two, as shown in Fig.~\ref{fig:simulation_scenario} ($\alpha = \beta = 2$).
\begin{figure*}
  \centering
     \subfigure[OTFS User.]{\includegraphics[width=0.32\textwidth]{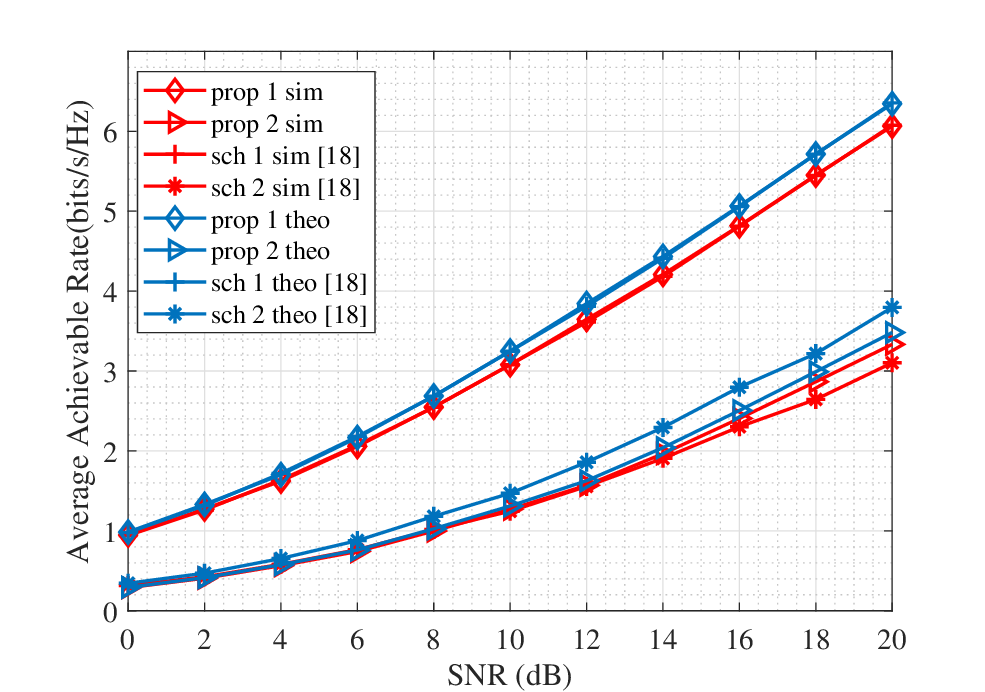}} 
     \subfigure[Worst User.]{\includegraphics[width=0.32\textwidth]{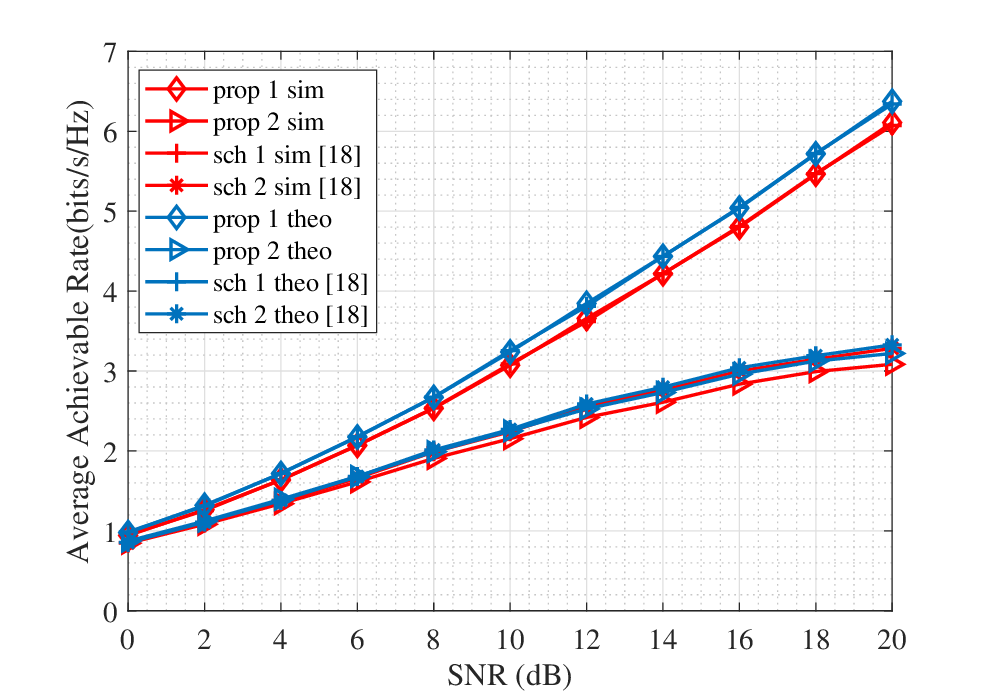}}
    \subfigure[Arbitrary User.]{\includegraphics[width=0.32\textwidth]{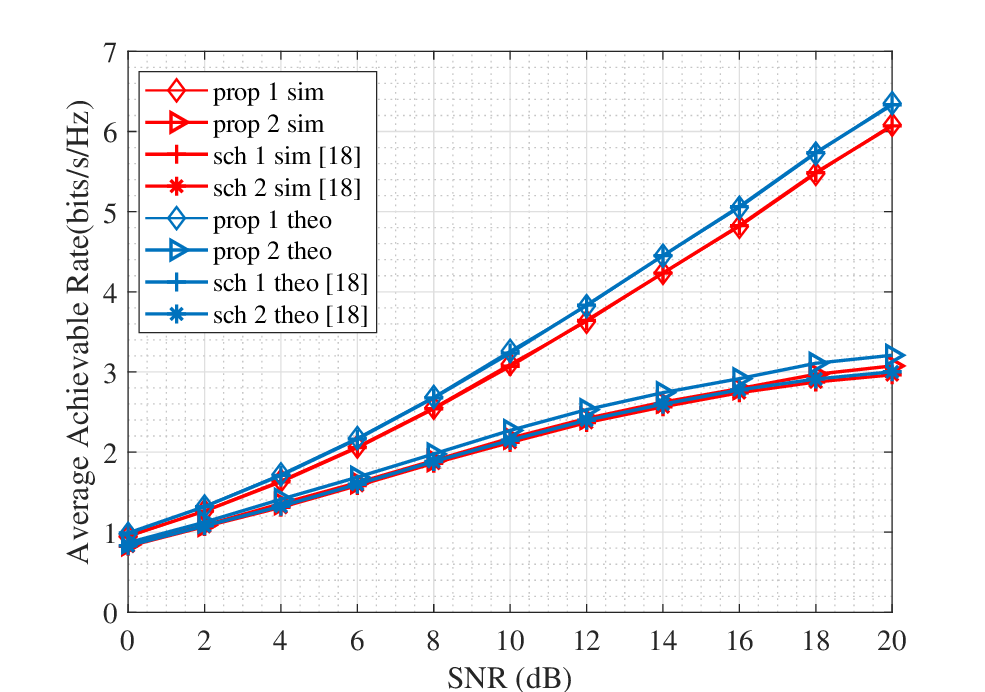}}
  \caption{The achievable rates for different schemes.}   
      \label{fig:rate_uplink1}
 \end{figure*}
 
\begin{figure}[t]
\centering
\includegraphics[scale = .5]{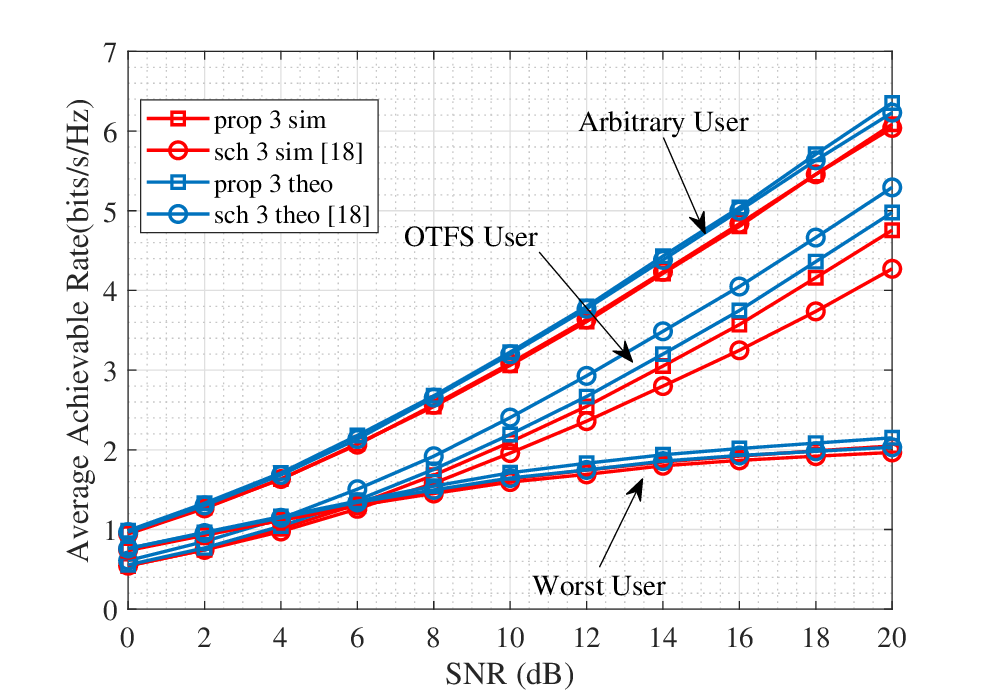}
\caption{The achievable rates for schemes ``prop 3'' and ``sch 3''.}
\label{fig:rate_uplink2}
\end{figure}

\setcounter{figure}{6}
\begin{figure}[t]
  \centering
     \subfigure[ The CCDF for different schemes.]{\includegraphics[scale=0.45]{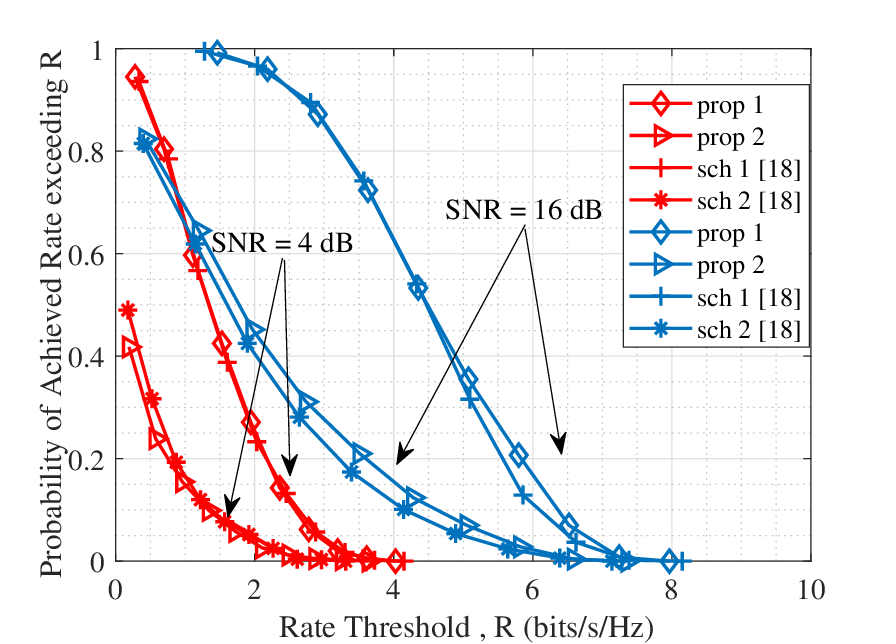}} 
     \subfigure[The CCDF for schemes `` prop 3'' and `` sch 3''.]{\includegraphics[scale=0.45]{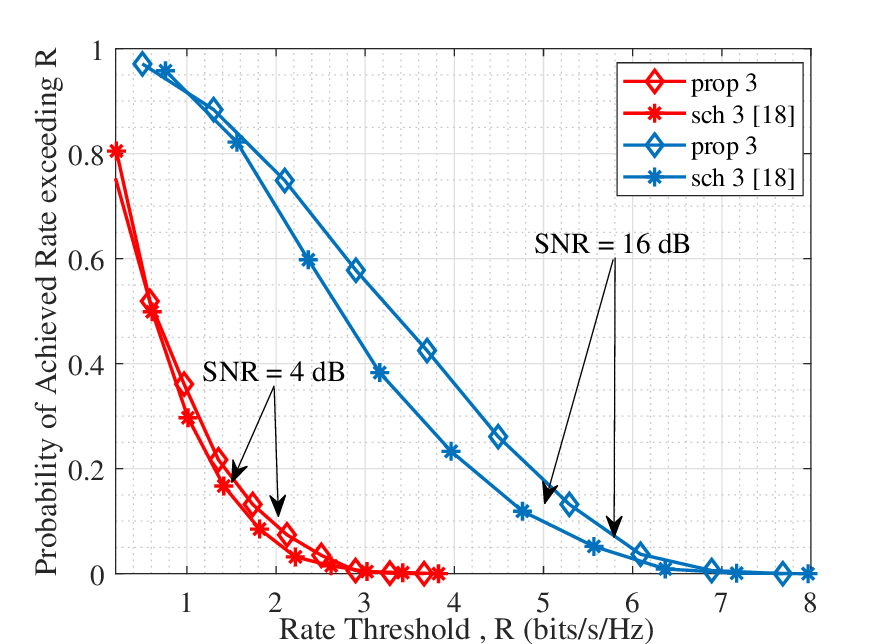}}
  \caption{The CCDF of the achieved rates for the OTFS user.}   
      \label{fig:ccdf}
 \end{figure}
 
\subsection{Achievable Rates}
This subsection examines achievable rates performance. Fig.~\ref{fig:rate_uplink1} and Fig.~\ref{fig:rate_uplink2} present the theoretical bounds (in blue) alongside the average achieved rates (in red). Each figure compares the proposed schemes—under different spreading parameter settings—with their unspread counterparts (e.g., see Fig. \ref{fig:simulation_scenario}), for the three types of users described earlier\footnote{For clarity, Fig. \ref{fig:rate_uplink1} shows results for ``prop 1'', ``sch 1'', ``prop 2'', and ``sch 2'', while Fig. \ref{fig:rate_uplink2} depicts ``prop 3'' and ``sch 3''.}.\\

Overall, the proposed schemes and their unspread counterparts exhibit comparable interference levels, resulting in similar upper bounds. However, the proposed schemes consistently show smaller gaps between the achieved rates and their respective bounds, indicating improved robustness. For example, at $\text{SNR} = 20~\text{dB}$, this gap for the OTFS User is around $1~\text{bits/s/Hz}$ for ``sch 3'', whereas it is improved to $0.2~\text{bits/s/Hz}$ with ``prop 3''. At the same $\text{SNR}$, this gap is improved from around $0.8~\text{bits/s/Hz}$ with ``sch 2'' to less than  $0.2~\text{bits/s/Hz}$ with ``prop 1''. Focusing on the proposed schemes, it is evident that the interference can be controlled by the proper selection of the spreading parameters. For instance, the Worst User's rate at an SNR of 20 dB can be increased from 3.2 bits/s/Hz to 6.2 bits/s/Hz by changing $\beta$ from $4$ to $1$, as shown in Fig.~\ref{fig:rate_uplink1}(b). For ``prop 1'', all users have similar performance, because all of them occupy different time indices, and as we showed by {\textbf{Lemma 3}}, this ensures no interference among them. However, the Worst User experiences more interference when $\beta$ is 2 (``prop 3'') than when $\beta =4$ (``prop 2''); the rate is increased from around 2 bits/s/Hz to 3 bits/s/Hz at 20 dB. The reason is that when $\beta=2$, this user suffers leakage from the OTFS User from both sides along the frequency dimension, but when $\beta = 4$, one side is 3 indices away from OTFS User. The OTFS User shows a constant trend: the achieved rate decreases for higher $\beta$ as depicted in Fig.~\ref{fig:rate_uplink1}(a). Even though there is no other OTFS User that would interfere with it, the SINR decreases with the increase in $\beta$, because as the OTFS User leaks more into other TF resources, it has less signal strength in its primary resources which degrades the achieved rates. The Worst User, Fig.~\ref{fig:rate_uplink1}(b), suffers most from interference, except when using the TDMA schemes. Finally, as shown in Fig.~\ref{fig:rate_uplink1}(c), the Arbitrary User achieves the highest rates in general, because it experiences the least amount of interference from the OTFS User.

\setcounter{figure}{7}
\begin{figure*}
  \centering
     \subfigure[OTFS User.]{\includegraphics[width=0.32\textwidth]{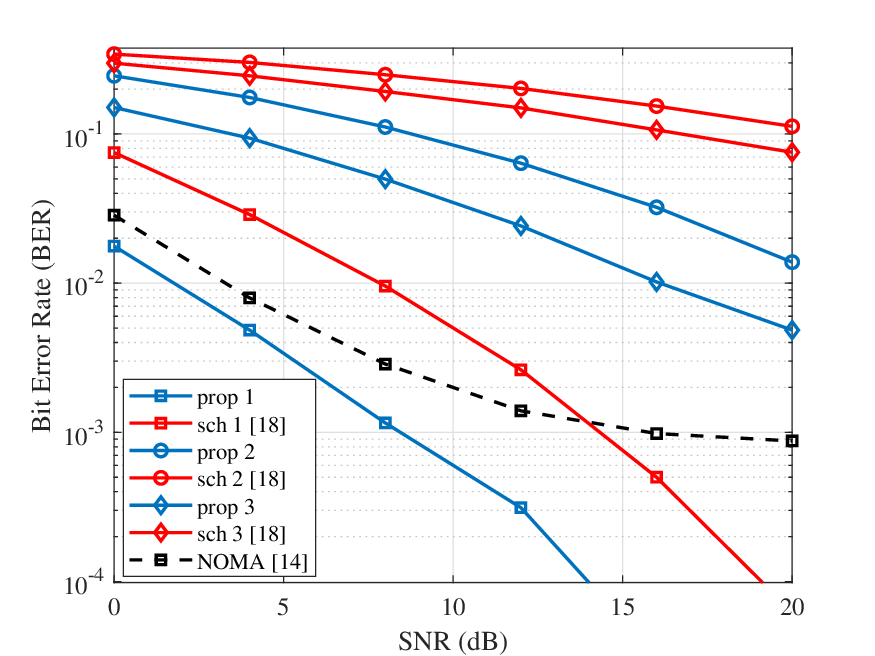}} 
     \subfigure[Worst User.]{\includegraphics[width=0.32\textwidth]{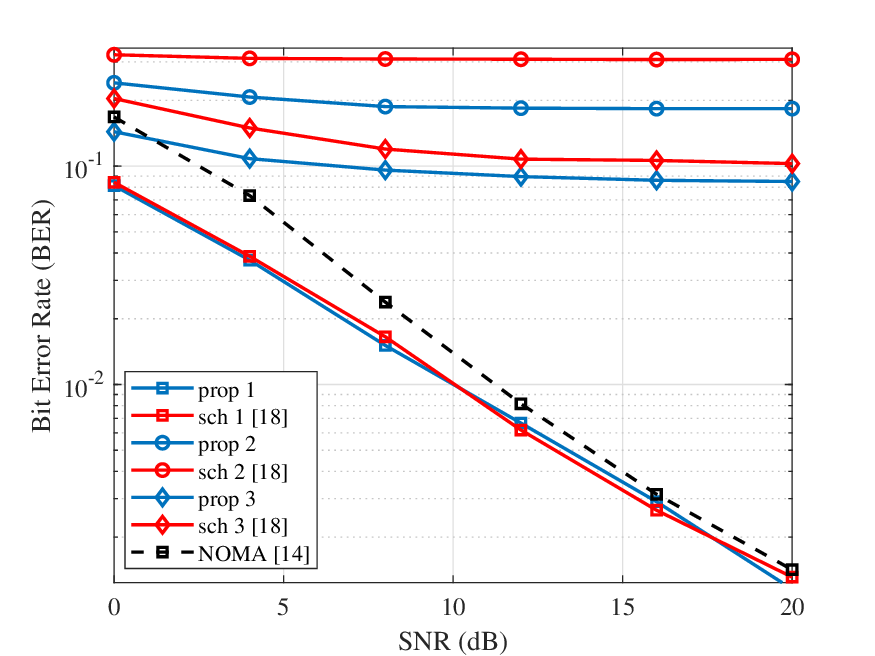}}
    \subfigure[Arbitrary User.]{\includegraphics[width=0.32\textwidth]{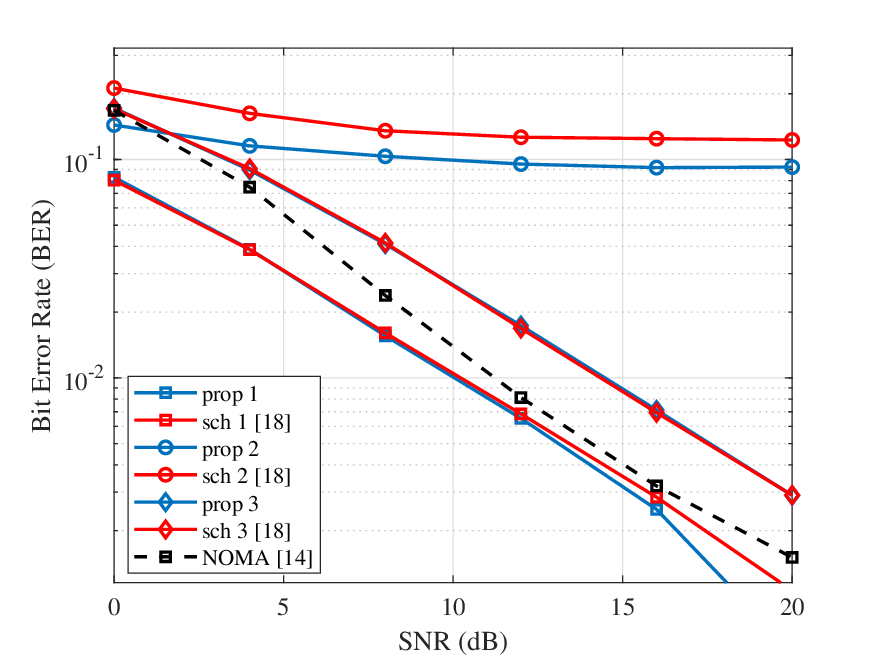}}
  \caption{BER for the three users using different coexistence schemes.}   
      \label{fig:BER_results}
 \end{figure*}

To provide a comprehensive view of the system's performance, we also present the \ac{CCDF} of the achieved rates for the OTFS User in Fig.~\ref{fig:ccdf}. Fig.~\ref{fig:ccdf}(a) shows the CCDF of the achievable rates of the OTFS User for ``prop 1'', ``sch 1'', ``prop 2'', and ``sch 2''; while Fig.~\ref{fig:ccdf}(b) shows the CCDF performance of ``prop 3'' and ``sch 3''. From the CCDF, it is evident that all proposed schemes show clear gains over their unspread alternatives. Notably, the CCDF ``prop 1'' is slightly shifted to the right, a trend that intensifies at higher SNR values.  
 
Observe the OTFS User's performance in Fig. \ref{fig:rate_uplink1}(a) and Fig. \ref{fig:rate_uplink2}. Except in TDMA schemes (prop 1'' and sch 1''), the OTFS User does not fully occupy the frequency domain and may leak outside its allocated resources, reducing signal strength in its own region. This leakage directly impacts the upper bounds from the rate analysis. Although both contiguous and interleaved allocations experience this effect, interleaved allocations tend to suffer more, as leakage can occur from any resource. In contrast, contiguous allocations mostly leak at the edges. Nonetheless, due to the TF diversity exploited by the interleaved design, average achievable rates may still improve despite slightly degraded upper bounds. 

In summary, ``prop 1'' achieves the maximum rates for all the users, and the interference is minimized. However, having freedom to choose the spreading parameters is advantageous to support a wide range of applications. While the proposed schemes exhibit similar interference patterns to their unspread counterparts, they demonstrate greater resilience and a superior ability to approach their upper bounds, due mainly to the greater diversity given to each user.

\subsection{Simulation of BER}

This subsection compares the BER performance of the three users under the various schemes defined above, as shown in Fig.~\ref{fig:BER_results}. It can be seen from Fig.~\ref{fig:BER_results}(b) and Fig.~\ref{fig:BER_results}(c) that the OFDM users exhibit similar trends across all schemes, with the interference being the main factor that determines the BER. However, for the Worst User, other than ``prop 1'' or ``sch 1'', the \ac{MUI} becomes overwhelming, leading to a substantial degradation in BER performance. In such cases, it is advisable to either switch to a different frequency resource or introduce guard bands to mitigate the interference. The same applies to the FDMA-based schemes, namely ``prop 2'' and ``sch 2'', where interference dominates the performance, and the BER shows no significant improvement at higher SNR values. This behavior arises because the frequency resources of the OTFS user cannot be completely avoided. Consequently, the use of a guard band becomes essential for the feasibility of these schemes. 


Focusing on the OTFS User, shown in Fig.~\ref{fig:BER_results}(a), ``prop 1'' and ``sch 1'' deliver the best BER performance, primarily due to their ability to minimize interference. The inherent advantage of the proposed scheme persists even under identical interference conditions, as ``prop 2'' and ``prop 3'' have about 12 dB gain over ``sch 2'' and `` sch 3'', which is the same gain for their lower bounds $3\alpha\beta~\text{dB}$, but it was not as significant for ``prop 1'', as other factors such as the interference dynamics and the gap between the average BER and the lower bound have to be taken into account. Nevertheless, the advantage of the proposed scheme can clearly be seen, and ``prop 1'' has over 6 dB gain compared to ``sch 1''.
From these simulations, the benefits of spreading in enhancing BER performance are evident, which is mainly due to the added diversity. However, the choice of spreading parameters must carefully consider the impact of interference to balance the trade-offs effectively.\\

Although NOMA is theoretically the most spectrally efficient, as it allows multiple users to reuse the same TF resources, this advantage comes at the cost of increased interference and higher receiver complexity, particularly due to the need for SIC. Therefore, a comprehensive comparison with NOMA would require a more nuanced analysis beyond the scope of this brief illustration.  Nevertheless, it is insightful to benchmark the BER performance of proposed coexistence scheme against NOMA-based systems.\\
Hence, in addition to the orthogonal schemes, we include the BER performance for a NOMA coexistence system, where OFDM users are orthogonally accessed in the TF domain, while an OTFS UE is spread across the entire TF plane according to a NOMA strategy. SIC is employed to recover the OFDM users' data symbols, following the approach described in \cite{ding2019otfs}. Notably, in this setting, all OFDM users achieve identical BER performance, as none is more susceptible to interference from the OTFS UE than the others. For the OTFS UE, we observe a clear gain in frequency diversity under the NOMA scheme, particularly at low SNR values where it outperforms the contiguous TDMA scheme. However, at high SNR values, the interference inherent to NOMA becomes more significant, reducing its advantage. The proposed scheme consistently outperforms the NOMA coexistence approach across all SNR levels, while for the OFDM users, SIC effectively suppresses OTFS interference, resulting in BER performance that approaches that of TDMA-based schemes. 

\subsection{Simulation of the Sensing Performance} 

This subsection compares the \ac{MSE} performance for the OTFS user with and without the \ac{TF} domain spreading. For more details about the evaluation, please refer to \cite{haif2024novel}.   
The results are shown in Fig.~\ref{fig:sensing_results}. In this setup, we consider a bistatic sensing configuration where the transmitter and sensing receiver are spatially separated. To isolate and emphasize the fundamental trade-off between resolution and unambiguous range/velocity, we assume that the sensing receiver has perfect knowledge of the transmitted waveform, i.e., a fully pilot-based signal structure. The advantage of the proposed system can be seen in the obtained performance. Other than achieving better resolution for both range and velocity, by a factor of $\alpha$ and $\beta$ respectively, the \ac{MSE} shows more robustness for high noise levels. For both range and velocity, the proposed system reaches the ideal resolution over 10 dB earlier than the non-spread scheme. In this case, since $\alpha = \beta = 8$, the resolution is improved by $10\text{log}(8)\approx 9.03~\text{dB}$, or almost tenfold, for both range and velocity.

\setcounter{figure}{8}
\begin{figure}[t]
\centering
\includegraphics[scale=.45]{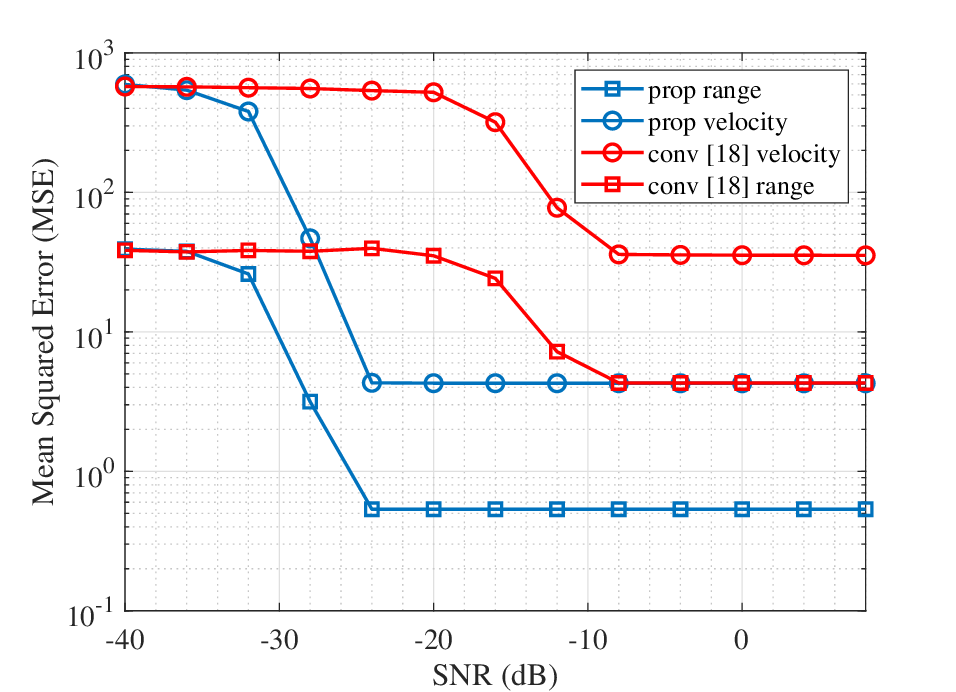}
\caption{Sensing performance for different schemes.}
\label{fig:sensing_results}
\end{figure}

\subsection{Effect of Spreading Parameters} 

As discussed earlier, $M'$ and $N'$ represent the nominal bandwidth and time duration that each user would occupy without spreading. After spreading, these dimensions expand by factors $\beta$ and $\alpha$, respectively, resulting in $M = M'\beta$ and $N = N'\alpha$ in the proposed scheme. While larger $\alpha$ and $\beta$ values are desirable for improving diversity and sensing resolution, increasing $\beta$ can introduce significant inter-user interference. Thus, it is preferable to maintain $\beta=1$ and increase $\alpha$ as much as is feasible, without violating the assumption of a quasi-static OFDM channel over the extended symbol duration. Therefore, the optimal choice of $\alpha$ and $\beta$ depends on the specific deployment scenario and system requirements.

\subsection{Effect of the number of users} 

In this subsection, we briefly examine how the number of users impacts the BER performance for the considered schemes. There are numerous possible allocation strategies for both the conventional and proposed designs, particularly as the number of users increases. Hence, it is impractical to explore all potential configurations exhaustively. Nevertheless, we focus on two representative cases that capture the two extrema of interference conditions, as any intermediate option would exhibit performance lying between these two extremes. Specifically, we simulate the effect on the OTFS UE for two orthogonal schemes—TDMA and FDMA—considering both the proposed and conventional approaches. Additionally, the performance of the NOMA-based coexistence scheme is presented for comparison.\\
Fig.~\ref{fig:different_numbers_of_users}(a) presents the BER performances for TDMA based schemes (``sch 1'' and ``prop 1''), along with NOMA. For ``prop 1'', we can see that the BER is improved for a greater number of users. This is due to the fact that in this scheme where TDMA is used, no leakage is present, and this remains true for any number of users. Hence, the number of users is represented only by $\alpha$, while $\beta = 1$: as we increase the number of uses, we increase the spreading, which enhances the BER performance as shown in the BER analysis. For the other TDMA scheme ``sch 1'', again no leakage is present. However, since diversity is not exploited through the spreading, the BER experiences little change as we increase the number of users. For the FDMA schemes (Fig.~\ref{fig:different_numbers_of_users}(b)), in spite of exploiting frequency diversity, leakage has a greater effect on the BER, which is degraded for higher numbers of users. This is true for both ``prop 2'' and ``sch 2'', but the proposed scheme still outperforms the conventional scheme for all numbers of users. We can extrapolate and state that for a combination of TDMA and FDMA in the proposed scheme, the BER enhances as more users are accessed using TDMA and is degraded as more users are accessed using FDMA. For the NOMA coexistence scheme, as expected, the BER worsens with an increasing number of users. Nonetheless, NOMA outperforms the conventional TDMA scheme at low SNR values and achieves better BER performance than the FDMA schemes across all SNR levels.

\setcounter{figure}{9}
\begin{figure}[t]
  \centering
     \subfigure[TDMA.]{\includegraphics[scale=0.45]{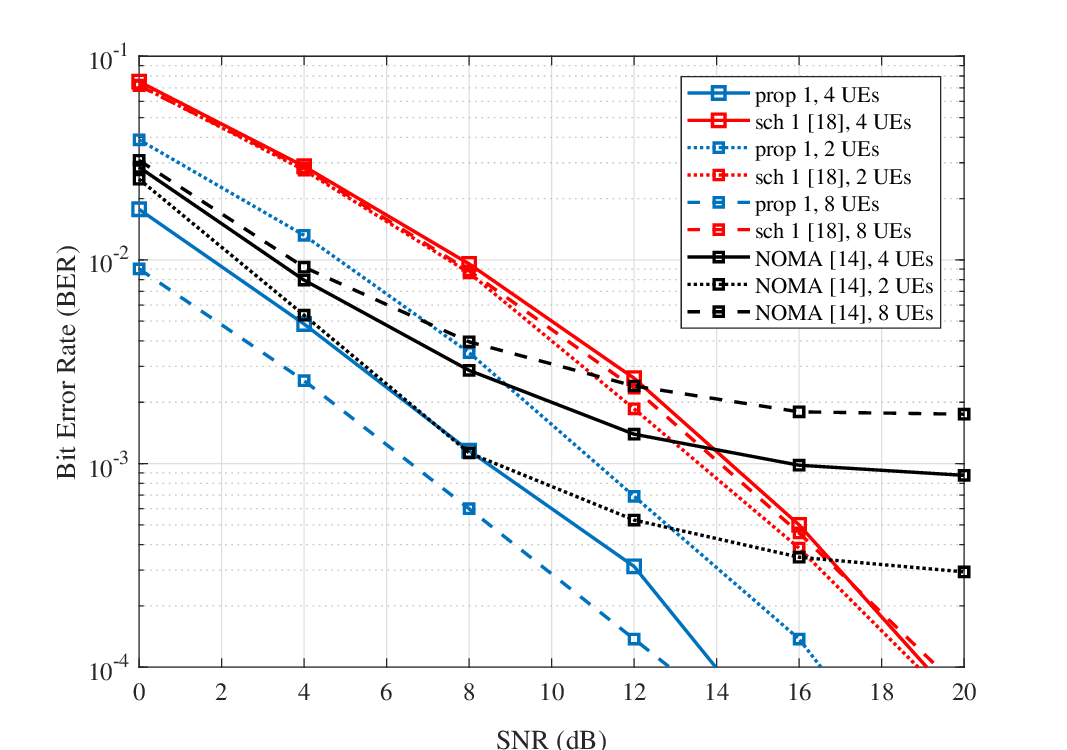}} 
     \subfigure[FDMA]{\includegraphics[scale=0.45]{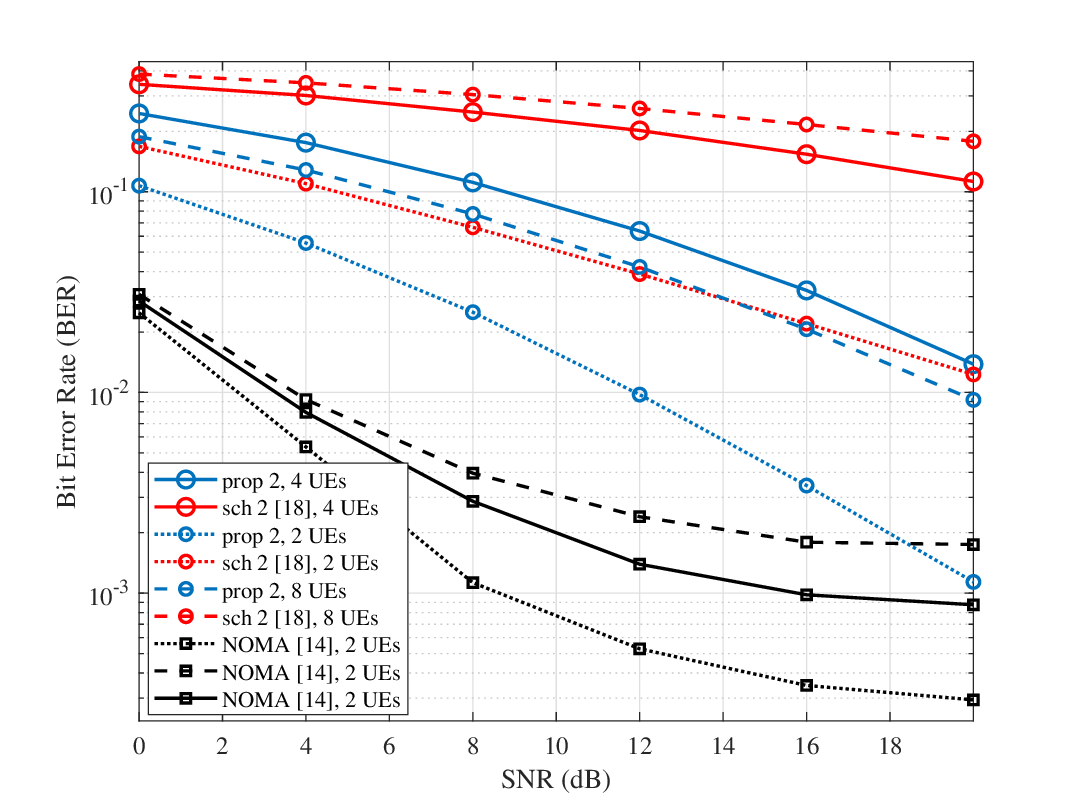}}
  \caption{BER performance for different numbers of users.}   
      \label{fig:different_numbers_of_users}
 \end{figure}

\section{Conclusion} 

In this paper, we introduced a novel coexistence scheme for OTFS and OFDM waveforms, leveraging a DD domain spreading approach to optimize resource utilization and minimize cross-interference. By employing a specifically designed spreading matrix, the proposed system ensures orthogonality between OTFS and OFDM, enabling effective coexistence within the same frequency band without significant degradation in performance.
The proposed scheme demonstrated several key advantages, including enhanced interference management, improved average achievable rates, and the ability to fully exploit channel diversity in multi-user, multi-waveform scenarios. Additionally, the system significantly improves range and velocity resolutions for OTFS users by utilizing the full bandwidth and time duration, contributing to superior sensing performance.
These findings underline the potential of the proposed system as a forward-looking solution for integrating OTFS and OFDM waveforms in 6G networks. 
\section*{Acknowledgment}
This study was supported by The Scientific and Technological Research Council of Türkiye (TUBITAK) under the Grant Number 123E514. The authors thank TUBITAK for their supports.
\bibliographystyle{IEEEtran}
\bibliography{IEEEfull}
\vfill
\end{document}